\documentclass[11pt]{article}
\usepackage{amscd,amsmath,amstext,amsfonts,amsbsy,amssymb,amsthm}

\usepackage[margin = 2.5cm]{geometry}

\usepackage[width=.85\textwidth, font = small, labelfont = bf, labelsep = endash, textfont = it]{caption}
\usepackage[titletoc,title]{appendix}

\usepackage[pagewise, mathlines]{lineno}
\usepackage{lscape, graphicx}
\begin{document}

% Title of paper
\title{Hierarchical multinomial processing tree models for meta-analysis of diagnostic accuracy studies}

% List of authors, with corresponding author marked by asterisk
\author{ANNAMARIA GUOLO\\[4pt]
% Author addresses
\textit{Department of Statistical Sciences,
University of Padova,
Via Cesare Battisti, 241-243,
Italy}
\\[2pt]
% E-mail address for correspondence
{annamaria.guolo@unipd.it}}

% Running headers of paper:
\markboth%
% First field is the short list of authors
{A. Guolo}
% Second field is the short title of the paper
{Hierarchical multinomial tree models in meta-analysis}

\maketitle

% Add a footnote for the corresponding author if one has been
% identified in the author list

\begin{abstract}
{Meta-analysis represents a widely accepted approach for evaluating the accuracy of diagnostic tools in clinical and psychological investigations. This paper investigates the applicability of multinomial tree models recently suggested in the literature under a fixed-effects formulation for assessing the accuracy of binary classification tools. The model proposed in this paper extends previous results to a hierarchical structure accounting for the variability between the studies included in the meta-analysis. Interestingly, the resulting hierarchical multinomial tree model resembles the well-known bivariate random-effects model under an exact within-study distribution for the number of true positives and true negatives subjects, with the additional advantage of providing an estimate of the prevalences of disease from each study. The proposal is in line with a latent-trait approach, where inference is performed according to a frequentist point of view.
The applicability of the proposed model and its performance with respect to the approximate bivariate random-effects model based on normality assumptions commonly used in the literature is evaluated in a series of simulation studies. Methods are applied to a real meta-analysis about the accuracy of the confusion assessment method as delirium screening tool.}
\end{abstract}
\hspace{1cm}
{{\small KEYWORDS:}}{ Diagnostic test; Prevalence; Random-effects; Sensitivity; Specificity.}

\section{Introduction}
\label{sec1}
The evaluation of a patient's disease status or the early detection of a certain disorder in clinical and psychological investigations is often reached through very accurate instruments, which are typically expensive, time-consuming, or discomforting. As a consequence, their large scale application is not appealing and the need for simpler, inexpensive, but still accurate classification tools remains an aim of the research. 

The accuracy of new proposed classification or diagnostic tools is evaluated through the comparison to a reference test assumed to be unquestionable, also called gold standard. In the last years, meta-analysis of diagnostic studies has been widely accepted as an approach for the assessment of the accuracy of a diagnostic or screening test in identifying a patient's specific status, or, more generally, in distinguishing between diseased and nondiseased patients. A diagnostic study is commonly evaluated in terms of sensitivity, i.e., the conditional probability of testing positive in subjects classified as positive by the reference test, and specificity, i.e., the conditional probability of testing negative in subjects classified as negative by the reference test.
As an alternative, a diagnostic test is evaluated using a two-by-two table of agreement between the test results and the reference test results (e.g., Honest and Khan, 2002).

The accuracy of a novel diagnostic test is often assessed using meta-analysis methods (e.g., Jackson et al., 2011). Within this framework, the bivariate hierarchical model (Reitsma et al., 2005; Arends et al., 2008) is currently a well-established technique. It is preferable to the traditional approach based on separate analyses for sensitivity and specificity, which do not account for the correlation between the diagnostic measures of accuracy. In addition, the bivariate hierarchical model improves on the popular proposal in Littenberg and Moses (1993) and in Moses et al. (1993) to construct a summary receiver operating characteristic curve based on the regression of the difference between sensitivity and specificity on their sum, a solution which has been criticised for not providing reliable inferential conclusions (e.g., Rutter and Gatsonis, 2001 ; Arends et al., 2008).
The bivariate model has a hierarchical structure accounting for the within-study sampling variability and for the between-study variability arising from differences due, for example, to study design's characteristics. Likelihood inference in this framework is affected by several issues (e.g., Guolo, 2017; Takwoingi et al., 2017). Authors warn against the risk of unreliable conclusions when the sample size is small, as well as the risk of non-convergence of the optimisation algorithms. Computational obstacles, e.g., the need for numerical integration, reduce the appealing of the approach. The mentioned issues leave space to alternative solutions, as, for example, solutions relaxing likelihood assumptions and relying on simulation strategies (e.g., Guolo, 2017).

This paper investigates the applicability of multinomial tree models, starting from a recent proposal in Botella et al. (2013) within the psychological literature, to assess the accuracy of binary classification tools. In particular, in this paper an extension of the fixed-effects multinomial tree model in Botella et al. (2013) is proposed, which turns out into a hierarchical model accounting for between-study heterogeneity. Interestingly, the resulting hierarchical multinomial tree model resembles the bivariate random-effects model under the exact -- binomial -- distribution for the number of true positives and true negatives within each study included in the meta-analysis. Likelihood-based inference for this model takes advantage of a clear separation of the parameters associated to the prevalence of the disease and parameters associated to the diagnostic accuracy measures. The performance of the method is compared to that of the likelihood-based approach for the classical bivariate random-effects model under the normal approximation for transformation of study sensitivity and specificity.
 
The methods are compared under different scenarios, including increasing sample size and increasing correlation between sensitivity and specificity. Scenarios include transformations of sensitivity and specificity given by logit function, probit function, and cloglog function. 

%A feature of interest is the evaluation of the methods when the reference test is perfect or imperfect. Multinomial tree models in Botella et al. (2013) have been developed by considering the possibility of imperfect reference test, while standard bivariate approach in meta-analysis does not. The simulation study performed in the paper will investigate the capability of the likelihood-based solution and of SIMEX to deal with such a substantial violation of the underlying assumptions, with respect to multinomial tree models.

The applicability of the competing methods is also evaluated on a meta-analysis about the accuracy of the confusion assessment method as delirium screening tool (Shi et al., 2013).

%It is worth mentioning that the methodology underlying multinomial tree models developed in Botella et al. (2013) does not allow for a hierarchical structure, so it does not account for between-study variability. This feature makes the comparison between multinomial tree models and the bivariate hierarchical model embedded in the long-lasting debate between the use of fixed-effects and random-effects models for meta-analysis, more in general, see, for example, Borenstein et al. (2010).

\section{Methods}
\label{sec2}
Consider a meta-analysis of $n$ diagnostic accuracy studies. Each study $i$, $i=1, \ldots, n$, provides information about the number of true positives $TP_i$, true negatives $TN_i$, false positives $FP_i$ and false negatives $FN_i$, see Table~\ref{table:data}. Let $P_i$ be the number of total positives and let $N_i$ be the number of total negatives.
The estimates of sensitivity ($SE_i$) and specificity ($SP_i$) can be obtained from study $i$ as $\widehat{SE}_i= TP_i/P_i$ and $\widehat{SP}_i = TN_i/N_i$, respectively. A common evaluation of the accuracy of the studies is in terms of a real-line transformation $\eta_i = g\left(SE_i\right)$ and $\xi_i = g\left(SP_i\right)$, with $g(\cdot)$ usually chosen to be the logit transformation. In this case, $\eta_i = logit\left(SE_i\right)=\log\{SE_i/(1-SE_i)\}$ and $\xi_i = logit\left(SP_i\right)=\log\{SP_i/(1-SP_i)\}$. Other choices are possible, as the probit transformation and the cloglog transformation, although rarely adopted.
The estimates of $\eta_i$ and $\xi_i$ represented by $\hat \eta_i$ and $\hat \xi_i$, respectively, are obtained from the sample counterpart.

\begin{table}[htp]
\caption{Two-by-two table of data from the comparison between the test under evaluation and the reference standard.}
\begin{center}
\begin{tabular}{cccc}
& \multicolumn{2}{c}{Reference test}&\\
Test & Disease status & No disease status& \\
\hline
Positives & $TP_i$ &$FP_i$& \\
Negatives &$FN_i$ & $TN_i$& \\
\hline
& $P_i$ & $N_i$ & $n_i$
\end{tabular}
\end{center}
\label{table:data}
\end{table}%

\subsection{Likelihood-based approach}\label{sec:lik}
The original bivariate random-effects model for meta-analysis of diagnostic accuracy studies follows the formulation developed in Reitsma et al. (2005) and in Arends et al. (2008). The model has a hierarchical structure distinguishing the within-study level, describing variability inside each study included in the meta-analysis, and the between-study level accounting for the heterogeneity among the studies, as a consequence of different study designs' or patients' characteristics. 

In line with the traditional approach, consider sensitivity and specificity expressed according to a transformation from the $[0,1]$ interval to the real line and let $\eta_i$ and $\xi_i$, respectively, denote such transformation. Usually, the logit transformation is adopted. At the within-study level, a model is specified for the observed $\left(\hat \eta_i, \hat \xi_i\right)^\top$ conditionally on the true study specific $\left(\eta_i, \xi_i\right)^\top$. A computationally convenient formulation is based on the normal approximation
\begin{equation}\label{eqn:within}
\left(
\begin{array}{c}
\hat \eta_i \\
\hat \xi_i
\end{array}
\right)\Bigg | \left(
\begin{array}{c}
 \eta_i \\
 \xi_i
\end{array}
\right) \sim N \left( 
 \left(
\begin{array}{c}
 \eta_i \\
 \xi_i
\end{array}
\right),
\Gamma_i
\right),
\end{equation}
with known diagonal variance/covariance matrix $\Gamma_i$, characterized by non-zero entries estimated in each study given the independence between positive and negative subjects at the within-study level. In case of logit transformation, 
 $$\Gamma_i = \left(
\begin{array}{cc}
P_i^{-1}+ \left(P_i-TP_i\right)^{-1}& 0\\
0 & N_i^{-1}+ \left(N_i-TN_i\right)^{-1}
\end{array}
\right).
$$
At the between-study level, the random effects $\left(\eta_i, \xi_i\right)^\top$ follow a normal distribution, namely,
\begin{equation}\label{eqn:between}
\left(
\begin{array}{c}
 \eta_i \\
 \xi_i
\end{array}
\right) \sim N \left( 
\mu=\left(
\begin{array}{c}
 \bar\eta \\
 \bar\xi
\end{array}
\right),
\Sigma=\left(
\begin{array}{cc}
 \sigma^2_\eta & \rho \sigma_\eta \sigma_\xi\\
 \rho \sigma_\eta \sigma_\xi & \sigma^2_\xi
\end{array}
\right)
\right),
\end{equation}
where $\bar \eta$ and $\bar \xi$ are the means over the studies, $\sigma^2_\eta$ and $\sigma^2_\xi$ are the between-study variances and $\rho$ is the correlation between $\eta_i$ and $\xi_i$. %It is expected that $\rho > 0$, as sensitivity $SE_i$ and specificity $SP_i$ tend to be negatively correlated and, consequently, $\eta_i$ and $\xi_i$ tend to be positively correlated.
The combination of (\ref{eqn:within}) and (\ref{eqn:between}) gives rise to a normal-normal model, with marginal specification
$$
\left(
\begin{array}{c}
\hat \eta_i \\
\hat \xi_i
\end{array}
\right) \sim N \left( 
\left(
\begin{array}{c}
 \bar\eta \\
 \bar\xi
\end{array}
\right),
\Gamma_i + \Sigma%\left(\begin{array}{cc}
 %\sigma^2_\eta & \rho \sigma_\eta \sigma_\xi\\
 %\rho \sigma_\eta \sigma_\xi & \sigma^2_\xi
%\end{array}\right)
\right).
$$
The associated log-likelihood function for the whole parameter vector $\theta=\left(\bar \eta, \bar \xi, \sigma^2_\eta, \sigma^2_\xi, \rho\right)^\top$ is available in closed form and it can be conveniently computed using standard software. Despite the computational advantages of the approach, several studies in the literature have highlighted the drawbacks, mainly related to the risk of unreliable inferential conclusions with few or sparse data. See, e.g., Chu et al. (2006), Guolo (2017), Takwoingi et al. (2017).
An alternative within-study model specification considers the exact distribution of observed true positives and false positives as realisations of binomial variables, instead of approximating the estimated transformations $\hat \eta_i, \hat \xi_i$ through a normal distribution. See, e.g.,  Arends et al. (2008) and Hamza et al. (2008). The exact binomial specification for the true positives and false positives combined with the normal specification (\ref{eqn:between}) for the between-study level gives rise to a marginal generalised linear model, with no closed-form expression for the associated likelihood function. Numerical integration is needed for likelihood computation and convergence problems can arise, in terms of non-positive definite variance/covariance matrix or estimates of the parameters of the variance/covariance matrix on the boundary of the parameter space. Such a drawback is more relevant in case of small sample size (e.g., Chen et al., 2017; Guolo, 2017; Takwoingi et al., 2017).

\subsection{Multinomial processing tree models}
Multinomial tree models (MTMs) represent a popular class of models for categorical behavioral data widely used in psychological research as an instrument to investigate cognitive processes (Riefer and Batchelder, 1988; Batchelder and Riefer, 1999; Erdfelder et al., 2009).
MTM analysis of categorical data is based on the assumption that the sample frequencies observed for a set of responses follow a multinomial distribution. As a relevant feature of the approach, interest is not only on the probabilities associated to the sample frequencies, but also to the path, or latent processes, leading to a response or behaviour of the cognitive process.
Differently from classical modeling of categorical data, e.g., using log-linear models or logit models,  
MTMs are structured in way to reflect a cognitive process, represented by a sequence of processing stages,
each of them resulting in a response category. The path followed by the cognitive process is conveniently represented by a tree with a single root, where each branch is a sequence of potential cognitive stages ending with the response category. The probability associated to each category is given by the sum of the probabilities associated to the branches leading to the same response (Batchelder and Riefer, 1999). Traditionally, data are aggregated across subjects, and then analyzed under the assumption of independently and identically distribution. Hierarchical extensions of the multinomial processing tree model accounting for between-subjects heterogeneity are the latent-trait approach and the beta-multinomial processing tree approach, both introducing random effects associated to the subjects specific parameters, although under different distributional specification. See, for example, Heck et al. (2018). The latent-trait approach in Klauer (2010) considers a probit transformation of the random components at the population level, following a multivariate normal distribution, in this way explicitly incorporating correlation structures. Inference is then performed according to a Bayesian perspective. The beta-multinomial processing tree approach assumes that the random components follow independent beta distributions. Not modelling potential correlations makes the approach less attractive. See also Jobst et al. (2020).
Both the hierarchical extensions of the MTMs represent a valid alternative to the latent-class approach (Klauer, 2006; Stahl and Klauer, 2007), where discrete population-level distributions model the between-study variability, with substantial computational effort.

\begin{figure}[htbp]
\begin{center}
\includegraphics[width=3.5in]{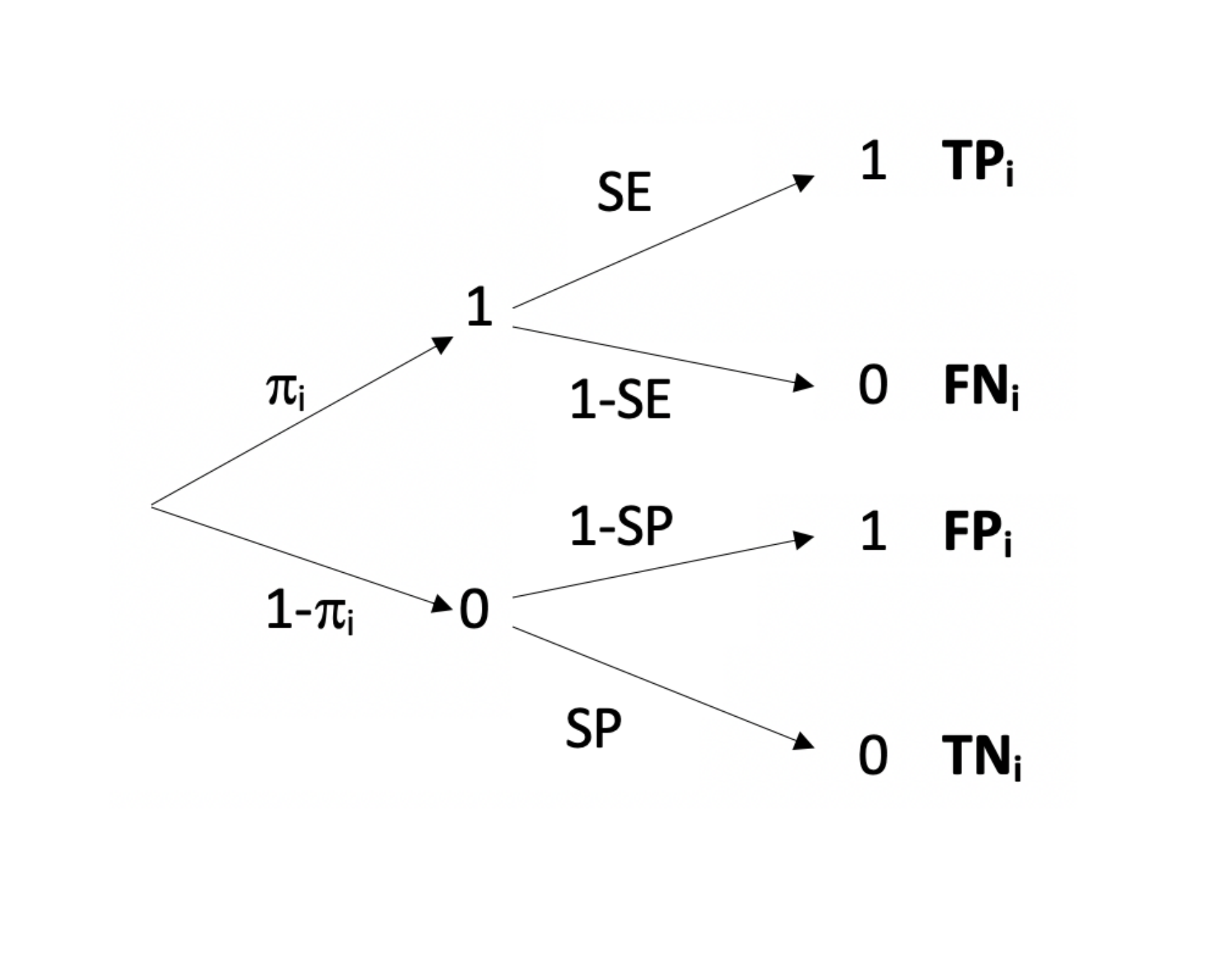}
\caption{Tree diagram associated to study $i$.}
\label{tree_perfect}
\end{center}
\end{figure}

The application of MTMs in meta-analysis of diagnostic accuracy studies has been originally proposed in Botella et al. (2013). According to this interpretation, the cognitive process is the result of the tools used to classify the patients according to their status, with response categories given by the cells in Table~\ref{table:data}.
Suppose that study $i$ included in the meta-analysis has an associated parameter $\pi_i$ reflecting the prevalence of the disease. Then, the tree diagram in Figure~\ref{tree_perfect} illustrates the process of assessment of the patients' status, under the assumption of perfect reference standard.

The multinomial tree model assumes that the studies are homogeneous and independent, with possibile different sample sizes and different prevalences $\pi_i$, $i=1, \ldots, n$, and common diagnostic measures $SE$ and $SP$. The probability associated to each response or category in study $i$ is
$$
p_{TP_i}=\pi_i SE; \hspace{1cm}
p_{FN_i}=\pi_i (1-SE); \hspace{1cm}
p_{FP_i}=(1-\pi_i) (1-SP);\hspace{1cm}
p_{TN_i}=(1-\pi_i) SP.
$$
The number of parameters to be estimated is $n+2$, with $2n-2$ degrees of freedom. As a consequence, the model needs at least two studies for estimation. 
The associated log-likelihood function for the whole parameter vector $\theta_{MTM}=\left(\pi_1, \ldots, \pi_n, SE, SP\right)^\top$ is 
\begin{eqnarray}\nonumber  
\ell(\theta_{MTM})&=& \sum_{i=1}^n \big[TP_i \log{\left(\pi_i SE\right)}  +  FN_i \log{\left\{\pi_i \left(1-SE\right)\right\}} + \\ \nonumber
&&FP_i \log{\left\{\left(1-\pi_i\right) \left(1-SP\right)\right\}}  +  TN_i \log{\left\{\left(1-\pi_i\right) SP\right\}}\big].
\end{eqnarray}
%%%%%%
% SPOSTARE IN FONDO. While meta-analysis of diagnostic accuracy studies is commonly developed under the assumption of error-free, or gold-standard, reference test, in practice the reference test is often imperfect. Reasons include the presence of errors and failures in the reference protocol or interpretation errors, non-existence of a gold standard, expensive costs of a gold standard. See Reitsma et al. (2009) for a detailed illustration. Using an imperfect reference standard as a perfect one can lead to unreliable conclusions about the accuracy of the diagnostic test as well as the prevalence of the disease, e.g., Valenstein (1990), Rutjes et al. (2006, 2007), Reitsma et al. (2009), Trikalinos and Balion (2012).

%A hierarchical extension of the multinomial processing tree model accounting for between-study heterogeneity follows a latent-class approach (Klauer, 2006; Stahl and Klauer, 2007) or a latent-trait approach (Klauer \cite{klauer2010}). In the first approach .... In the latent-trait approach, a parametric distribution is chosen for the parameters at the population level, with parameters estimated from the data. Ansari et al. \cite{ansari2008} suggest a multivariate normal distribution after a logit transformation of the parameters of test accuracy (MAI PUBBLICATO) , while Klauer \cite{klauer2010} employs a probit link. Both the approaches carry out inference according to a Bayesian perspective.

In this paper we consider a hierarchical extension of the multinomial tree model in Botella et al. (2013) for meta-analysis of diagnostic tests. Let $g(\cdot)$ denote a general link function to translate the test accuracy measure $SE$ and $SP$ to the real line and use a multivariate normal distribution to model the random-effects associated to the transformation of $SE$ and $SP$. The link function $g(\cdot)$ is not restricted to probit, as in Klauer (2010), but it can be chosen among classical transformations as logit, probit, cloglog. See, for example, Chen et al. (2017) for an evaluation of the composite likelihood approach for meta-analysis of diagnostic accuracy studies under different link functions. Differently from the latent-trait approach in Klauer (2010), inference will be performed from a frequentist point of view.
%In this way, the between-study level will be similar to the one used in the classical likelihood-based approach to meta-analysis of diagnostic test accuracy studies, as described in Section~\ref{sec:lik}. Differently from Ansari et al. \cite{ansari2008}, inference will be performed from a frequentist point of view, and the presence of study-specific prevalences inside the vector of parameters to be estimated will be properly taken into account.
Let the $g(\cdot)$ link function be
$$
g\left(SE_i\right)=\eta_i=\log\frac{SE_i}{1-SE_i}, \ \ g\left(SP_i\right)=\xi_i=\log\frac{SP_i}{1-SP_i}
$$
in case of logit transformation, 
$$
g\left(SE_i\right)=\eta_i= \Phi^{-1}\left(SE_i\right), \ \ g\left(SP_i\right)=\xi_i =\Phi^{-1}\left(SP_i\right)
$$
in case of probit transformation, 
$$
g\left(SE_i\right)=\eta_i= \log\left\{-\log\left(1-SE_i\right)\right\}, \ \ g\left(SP_i\right)=\xi_i =\log\left\{-\log\left(1-SP_i\right)\right\}
$$
in case of cloglog transformation. Then, given the above specifications, the hierarchical version of the MTM model in Botella et al. (2013) has associated log-likelihood function for the whole parameter vector $\theta_{MTM}=\left(\pi_1, \cdots, \pi_n, \bar \eta, \bar \xi, \sigma^2_\eta, \sigma^2_\xi, \rho\right)^\top$ equal to 
\begin{equation}\label{eqn:likMTM}
\ell_{MTM}\left(\theta_{MTM}\right)= \sum^n_{i=1} \log \int \int \pi^{P_i} \left(1-\pi\right)^{N_i} \eta_i^{TP_i} \left(1-\eta_i\right)^{FN_i} \left(1-\xi_i\right)^{FP_i} \xi_i^{TN_i} \phi_2\left(\eta_i,\xi_i; \mu; \Sigma\right) d\eta_i d\xi_i,
\end{equation}
where $\phi_2\left(\eta_i, \xi_i; \mu; \Sigma\right)$ is the density function of the bivariate normal distribution for $\left(\eta_i, \xi_i\right)^\top$ with mean $\mu$ and variance/covariance matrix $\Sigma$, as in (\ref{eqn:between}).
Consider that the log-likelihood function (\ref{eqn:likMTM}) has separable parameters, with nuisance components associated to the disease study prevalences $\left(\pi_1, \cdots, \pi_n\right)^\top$ well separated from the diagnostic accuracy parameters $\left(\bar \eta, \bar \xi, \sigma^2_\eta, \sigma^2_\xi, \rho\right)^\top$, so that 
\begin{equation}\label{eqn:likMTM2}
\ell_{MTM}\left(\theta_{MTM}\right)= \ell_{MTM,1} \left(\pi_1, \cdots, \pi_n\right) + \ell_{MTM,2} \left(\bar \eta, \bar \xi, \sigma^2_\eta, \sigma^2_\xi, \rho\right).
\end{equation}
The separability of the parameters in the likelihood function implies that inference on the diagnostic accuracy parameters can be based only on
\begin{equation}\label{eqn:likMTMparte2}
\ell_{MTM,2} \left(\bar \eta, \bar \xi, \sigma^2_\eta, \sigma^2_\xi, \rho\right)=\sum^n_{i=1} \log \int \int \eta_i^{TP_i} \left(1-\eta_i\right)^{FN_i} \left(1-\xi_i\right)^{FP_i} \xi_i^{TN_i} \phi_2\left(\eta_i,\xi_i; \mu; \Sigma\right) d\eta_i d\xi_i.
\end{equation}
%Differently from the exact likelihood approach, the use of the hierarchical MTM approach allows an ease estimation of study-specific disease prevalences, based on the reduced likelihood
Whichever the specification of the link function $g(\cdot)$, the two-dimensional integral needs to be solved numerically, for example, via Gauss-Hermite quadrature. Interestingly, the log-likelihood function (\ref{eqn:likMTMparte2}) resembles the log-likelihood function obtained by substituting the within-study approximate distribution (\ref{eqn:within}) with the exact distribution of $TP_i$ and $TN_i$, given by the binomial variables 
$$
TP_i \sim Binom\left(P_i, \eta_i\right), \ \ TN_i \sim Binom\left(N_i, \xi_i\right),
$$
as illustrated in Guolo (2017), among others. Differently from the exact likelihood approach, the use of the hierarchical MTM approach allows to estimate the within-study prevalences $\pi_i, i=1, \ldots, n$. Although the presence of prevalences $\pi_i, i=1, \ldots, n$ makes the dimension of the whole parameter vector increase with the sample size, the special structure of the likelihood function (\ref{eqn:likMTM2}) with separable parameters allows a straightforward independent estimation of the prevalences, based on the restricted likelihood
$$
\ell_{MTM,1} \left(\pi_1, \cdots, \pi_n\right) =\sum^n_{i=1} \log \pi^{P_i} \left(1-\pi\right)^{N_i}.
$$
The estimate of the study-specific disease prevalences can be obtained in closed form,
$$
\hat \pi_i = \frac{P_i}{n_i}
$$
as the fraction of positives in each study of dimension $n_i$, 
with standard errors given by
$$
se(\hat \pi_i)= n_i^3 \left(P_i^{-1} \cdot N_i^{-1} \right).
$$
With reference to the diagnostic accuracy parameters, instead, an appropriate evaluation of standard error is via the sandwich method, see Kauermann and Carroll (2001).

 %an ease estimation of study-specific disease prevalences, based on the reduced likelihood. Note that the presence of $\pi_i, i=1, \ldots, n$ makes the dimension of the parameter vector increase with the sample size, that is an example of the Neymann-Scott problem RIF and CONSEQUENCES. The potential issues related to the presence of $\pi_i$ can be faced using a two-step pseudo-likelihood approach, as in Gong and Samaniego (). Let $\hat \pi_i$ be an estimate of $\pi_i$ obtained in a simplified way, other than by maximizing   $\ell_{MTM}(\theta_{MTM})$. Then, parameters $\pi_i$ in (\ref{eqn:likMTM}) are assumed to be known and fixe to the estimated value $\hat \pi_i$. The resulting pseudo-likelihood to be maximized is
%\begin{equation}\label{eqn:likMTM2}
%\ell_{MTM2}(\theta_{MTM})= \sum^n_{i=2} \log \int \int \hat \pi^{P_i} (1-\hat\pi)^{N_i} \eta_i^{TP_i} (1-\eta_i)^{FN_i} (1-\xi_i)^{FP_i} \xi_i^{TN_i} \phi_2() d\eta_i d\xi_i.
%\end{equation}
%with parameter vector $\theta_{MTM2}=\left(\bar \eta, \bar \xi, \sigma^2_\eta, \sigma^2_\xi, \rho\right)^\top$. The standard error of the maximum pseudo-likelihood estimator of $\theta_{MTM2}$ is obtained by sandwich method, as...
%$$
%rl()
%$$
%The estimate of the study-specific disease prevalences can be obtained in closed form as .....
%$$
%\hat \pi_i = 
%$$
\section{Simulation study}
The performance of the proposed hierarchical multinomial tree model has been investigated through a series of simulation studies under a variety of scenarios and compared to that of the likelihood-based approach for the bivariate random-effects model under the normal approximation for transformation of study sensitivity and specificity.
Data simulation follows a two-stage procedure. First, for given number of studies $n$, the sample size $n_i$ of each study included in the meta-analysis is generated from a uniform variable on $[50, 200]$ and the number of true positives in each study is generated from a binomial distribution with parameters $n_i$ and a given prevalence of the disease. Then, for each true positive or true negative in the study, the corresponding classification provided by the test under evaluation is obtained as the result of a binomial distribution with probability of success given by the inverse of the link function $g(\cdot)$. We distinguish logit function, probit function, and cloglog function. The comparison between the data generated at the first step and the data generated at the second step provides the two-by-two Table~\ref{table:data} for study $i$. From the Table, quantities $\hat \eta_i$ and $\hat \xi_i$ can be determined according to the chosen link function $g(\cdot)$. Examined scenarios include sample size $n$ varying in $\{10, 25\}$, increasing prevalence of the disease at the population level, ranging in $\{0.20, 0.35\}$, increasing correlation between accuracy measures $\rho \in \{0.2, 0.6, 0.8\}$, and large accuracy of the test $\left(SE, SP\right)^\top=\left(0.9, 0.85\right)^\top$ or smaller accuracy of the test $\left(SE, SP\right)^\top=\left(0.80, 0.92\right)^\top$. The simulation is based on 1,000 replicates of each scenario. All the methods are implemented in the \texttt R programming language (R Core Team, 2022). The code is is available at https://github.com/annamariaguolo/MTM-meta-analysis.

Maximum likelihood estimation is carried out using the Nelder and Mead algorithm (Nelder and Mead, 1965), with integral evaluation for the MTM model based on the Gauss-Hermite quadrature with 21 nodes. Starting values for the optimization procedure are given by the empirical estimates of sensitivity and specificity and the empirical prevalences for the MTM model. In case of nonconvergence of the optimization algorithm, other solutions have examined, as the quasi-Newton BFGS algorithm and changes of the starting values of the parameters. 

\subsection{Results}
Methods are compared in terms of bias, standard deviation, and average of standard errors of the estimators of the parameters. Empirical coverages for Wald-type confidence intervals at nominal level 0.95 for parameters $\bar\eta$ and $\bar\xi$ are also reported. Results under nonconvergence were excluded when evaluating the simulations results. The failure rate of the approaches is another criterion used for comparison. 

\begin{figure}[!p]
\centering
\includegraphics[width=5in]{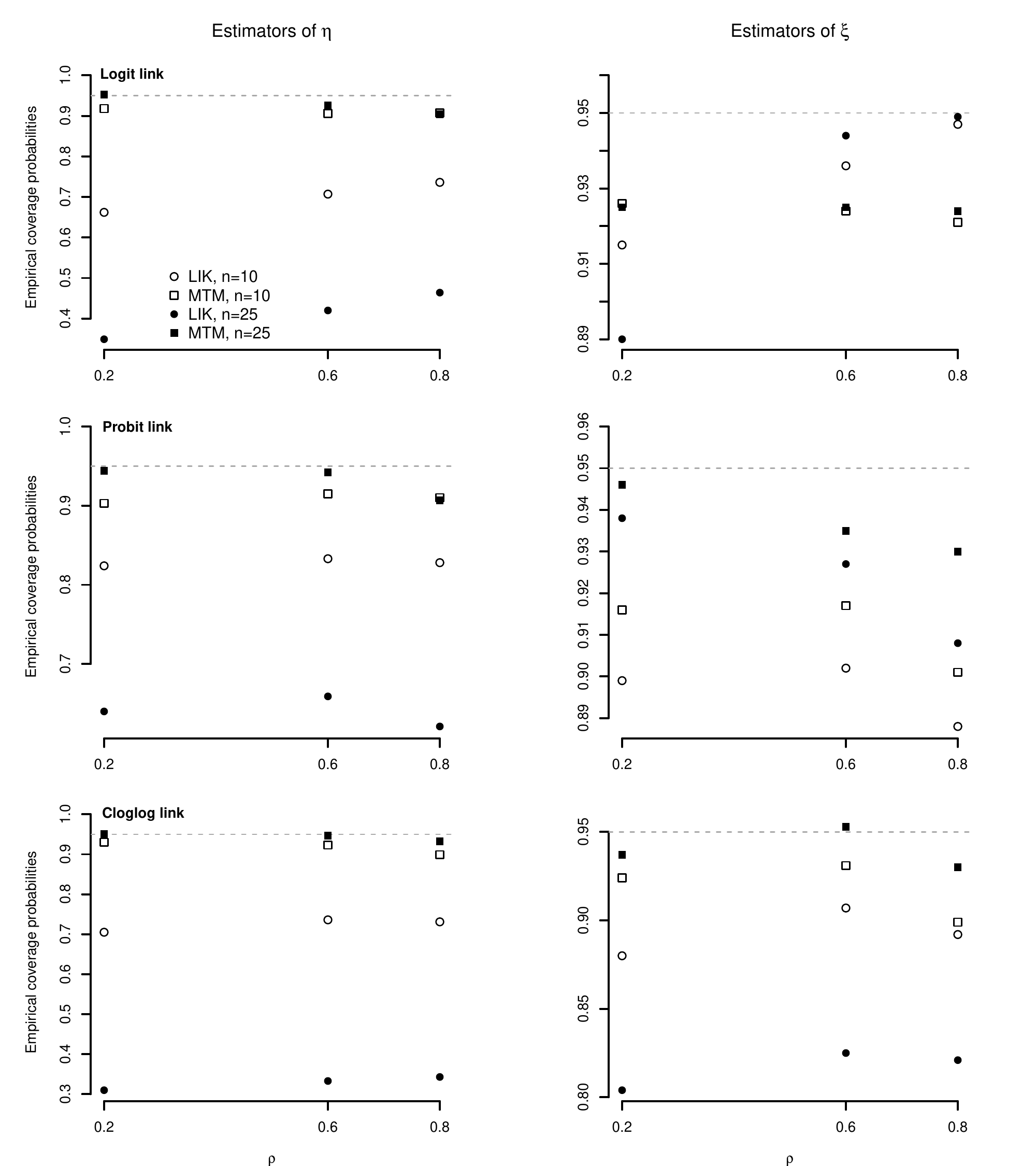}
\caption{{ Empirical coverage probability of Wald-type confidence interval for the estimators of $\bar\eta$ and $\bar\xi$ obtained from the approximate likelihood approach (LIK) and from the MTM approach (MTM), under increasing $\rho$, increasing sample size $n$ and different link function. High accuracy of the test. Prevalence of disease $\pi=0.20$. Dashed line: nominal 95\% level.}}
\label{fig:cov_pi20}
\end{figure}

\begin{figure}[htbp]
\begin{center}
\includegraphics[width=5in]{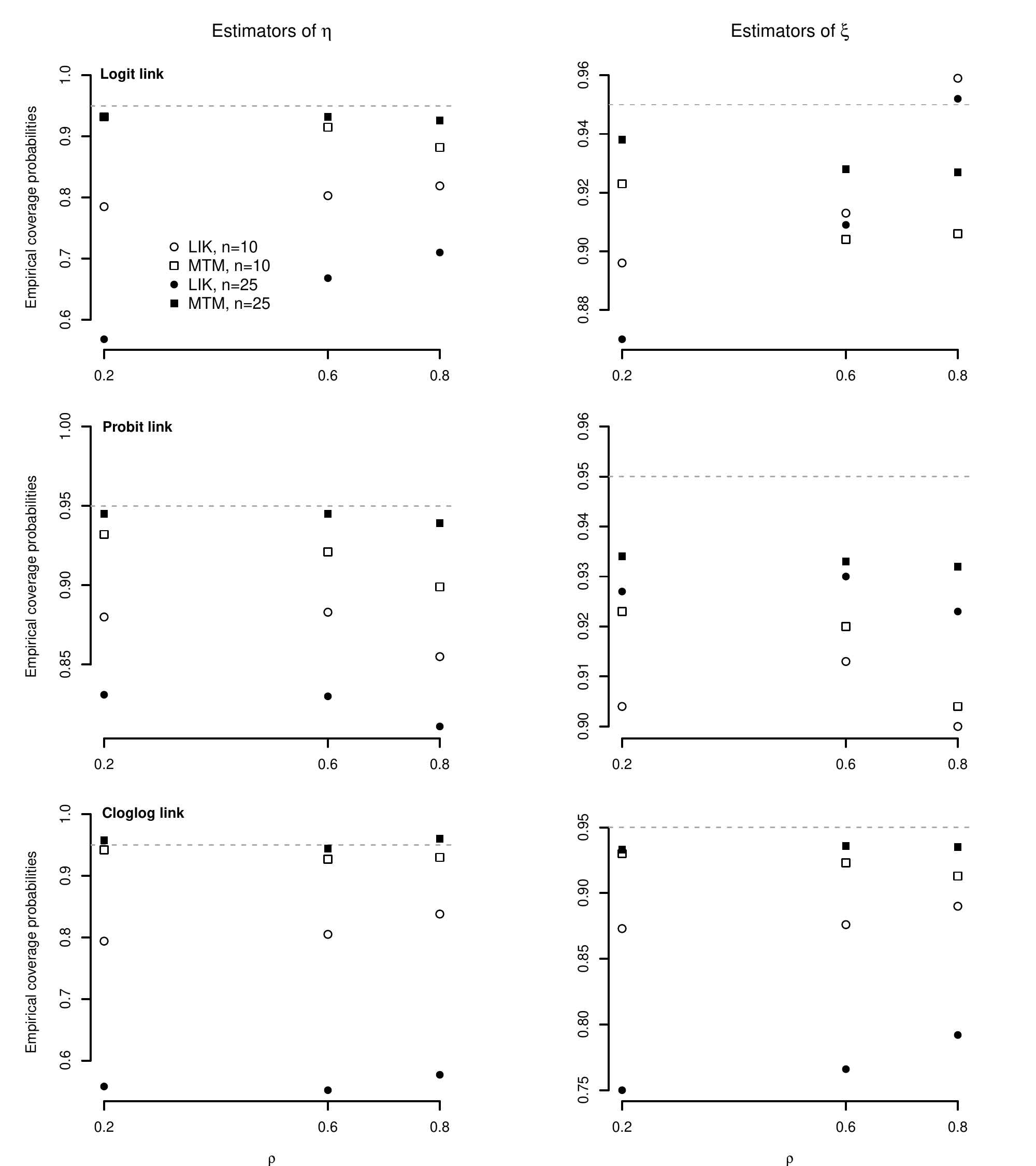}
\caption{{ Empirical coverage probability of Wald-type confidence interval for the estimators of $\bar\eta$ and $\bar\xi$ obtained from the approximate likelihood approach (LIK) and from the MTM approach (MTM), under increasing $\rho$, increasing sample size $n$ and different link function. High accuracy of the test. Prevalence of disease $\pi=0.35$. Dashed line: nominal 95\% level.}}
\label{fig:cov_pi35}
\end{center}
\end{figure}

Tables~\ref{tab:1}-\ref{tab:2} report the bias, the standard deviation and the average of standard error for the estimators of the fixed-effects components $\bar\eta$ and $\bar\xi$, for the estimators of the variance components $\sigma^2_\eta$ and $\sigma^2_\xi$, and for the estimator of the correlation parameter $\rho$, under increasing values of $\rho$, different link functions $g(\cdot)$, large accuracy of the test $\left(SE, SP\right)^\top=\left(0.9, 0.85\right)^\top$, small number of studies included in the meta-analysis $n=10$, by distinguishing small and large prevalence of the disease, equal to 0.20 and to 0.35, respectively. Similar results for low accuracy of the test and for larger sample size are reported in the Supplementary Material.
The use of the approximate likelihood approach gives rise to more biased estimates of the parameters, especially those related to $\bar\eta$ and the variance components, if compared to the MTM solutions. The correlation parameter tends to be overestimated. Such a behaviour of the approximation likelihood approach is more evident in case of logit link, and for increasing values of the correlation $\rho$, with biased more pronounced under low prevalence of the disease, see Table~\ref{tab:1}. Results tend to be less biased under the probit link. The use of MTM, conversely, produces almost unbiased estimators, without being substantially affected by changes of the link function, of the values of the correlation or of the prevalence of the disease. The price to pay is a slight increase of the variability associated to the estimates. 
Such a relative behaviour of the competing methods is maintained under increasing sample size $n=25$, see the corresponding results in Tables S1-S2 in the Supplementary Material. For both the approaches, increasing the sample size gives rise to a better performance in terms of bias of the estimators of the fixed-effects parameters, the variance components and the correlations, and in terms of the associated variability, as expected.
Figure~\ref{fig:cov_pi20} reports the empirical coverage of Wald-type confidence intervals at nominal level 95\% for the estimators of $\bar\eta$ and $\bar\xi$ under small and large sample size, for low disease prevalence $\pi=0.20$.
The advantages of MTM in reaching the target 95\% si substantial. The use of the approximate likelihood approach provides empirical coverage probabilities notably below the target level, as a consequence of biased estimates of the parameters, whichever the link function. Increasing the sample size does not help, and results can be even worse, as it happens with reference to the estimator of $\bar\eta$.
The use of MTM provides empirical coverage probabilities closer to the 95\% level, with an improved performance as the sample size increases.
Similar results are obtained for larger prevalence of disease, $\pi=0.35$, see Figure~\ref{fig:cov_pi35}.

Methods behave substantially differently from a computational point of view. The approximate likelihood does not require computational effort, while the application of the MTM approach requires numerical integration. This leads also to difference in terms of failure rate. Failure is a consequence of estimates of the correlation parameters on the boundary of the parameter space and/or not positive definite variance/covariance matrix and it is mainly experienced in case of small sample size $n=10$. Convergence problems affect both the approaches, and the MTM solution in particular, in case of logit link and small value of the correlation parameter $\rho$. The result is not unexpected, as previous findings in the literature of meta-analysis of diagnostic accuracy studies confirm the risk of convergence problems for the exact likelihood approach. See, for example, Guolo (2017),   Takwoingi et al. (2017), and Chen et al. (2017). The failure rate of the approximate likelihood solution tends to deeply reduce under the probit link or the cloglog link, while the decay for the MTM approach is slower. Increasing the sample size helps to eliminate the convergence problems for both the approaches.

When the accuracy of the test is low, namely, $\left(SE, SP\right)^\top=\left(0.80, 0.92\right)^\top$, the bias is reduced for all the estimators of the parameters, especially under the approximate likelihood solution. See the results reported in Tables S1-S2 for $n=10$ and Tables S3-S4 for $n=25$, both when the prevalence of the disease is equal to $0.20$ and $0.35$. The empirical coverages probabilities at nominal level 0.95 confirm a satisfactory behaviour of MTM under all the examined scenario, with values closer to the target level than the likelihood approach, although differences between the approaches are less marked than under the large accuracy case. See Figures S1-S2 in the Supplementary Material. Convergence problems are reduced for both the approaches, if compared to the high accuracy case.

\begin{landscape}

\begin{table}[hbtp]
\caption{Bias (standard deviation s.d., average standard error s.e.) for the estimators of $\bar\eta, \bar\xi, \sigma^2_\eta, \sigma^2_\xi, \rho$ obtained from the approximate likelihood approach (LIK) and from the MTM approach (MTM), under increasing $\rho$ and different link function. High accuracy of the test. Sample size $n=10$. Prevalence of disease $\pi=0.20$.}
\begin{center}
\begin{tabular}{c c c c c c c}
Method & $\rho$ & $\bar\eta$ & $\bar\xi$& $\sigma^2_\eta$& $\sigma^2_\xi$&  $\rho$\\
&& Bias (s.d., s.e.) & Bias (s.d., s.e.) & Bias (s.d., s.e.) & Bias (s.d., s.e.)& Bias (s.d., s.e.) \\
\hline
{\it Logit link} &&&&&&\\
LIK & 0.2 & -0.622 (0.313, 0.380) &  -0.054 (0.237, 0.235) & -0.680 (0.436, 0.342) & -0.123 (0.244, 0.206) & 0.066 (0.347, 0.252) \\ 
MTM && -0.029 (0.465, 0.461) & 0.022 (0.268, 0.259) & -0.290 (0.697, 0.586) & -0.034 (0.319, 0.268)  & 0.009 (0.583, 0.399)\\

LIK & 0.6 & -0.574 (0.320, 0.371) &  -0.020 (0.232, 0.243) & -0.664 (0.414, 0.339) & -0.061 (0.313, 0.234) & 0.190 (0.291, 0.218) \\ 
MTM && -0.021 (0.477, 0.459) & 0.016 (0.263, 0.270) & -0.199 (0.674, 0.603) & -0.004 (0.398, 0.290)  & 0.000 (0.481, 0.304)\\

LIK & 0.8 & -0.551 (0.312, 0.367) &  0.024 (0.237, 0.251) & -0.647 (0.434, 0.349) & -0.019 (0.320, 0.258) & 0.238 (0.244, 0.180) \\ 
MTM && 0.026 (0.467, 0.445) & 0.032 (0.274, 0.275) & -0.154 (0.658, 0.627) & 0.033 (0.371, 0.318)  & 0.024 (0.363, 0.235)\\

{\it Probit link} &&&&&&\\
LIK & 0.2 & -0.243 (0.267, 0.236) &  -0.024 (0.207, 0.193) & -0.697 (0.327, 0.130) & -0.131 (0.165, 0.130) & 0.049 (0.324, 0.241) \\ 
MTM && -0.001 (0.426, 0.415) & 0.015 (0.261, 0.257) & -0.102 (0.768, 0.582) & -0.018 (0.333, 0.268)  & -0.002 (0.463, 0.320)\\

LIK & 0.6 & -0.250 (0.258, 0.240) &  -0.038 (0.203, 0.192) & -0.679 (0.334, 0.240) & -0.135 (0.169, 0.131) & 0.156 (0.251, 0.196) \\ 
MTM && -0.014 (0.392, 0.408) & 0.002 (0.255, 0.249) & -0.097 (0.769, 0.581) & -0.029 (0.315, 0.248)  & -0.029 (0.345, 0.226)\\

LIK & 0.8 & -0.244 (0.275, 0.240) &  -0.043 (0.218, 0.195) & -0.684 (0.341, 0.239) & -0.134 (0.154, 0.132) & 0.202 (0.193, 0.155) \\ 
MTM && -0.003 (0.418, 0.391) & 0.008 (0.279, 0.258) & -0.138 (0.741, 0.521) & -0.017 (0.300, 0.247)  & -0.034 (0.224, 0.125)\\

{\it Cloglog link} &&&&&&\\
LIK & 0.2 & -0.344 (0.219, 0.211) &  -0.138 (0.174, 0.165) & -0.831 (0.234, 0.202) & -0.235 (0.147, 0.109) & 0.037 (0.313, 0.230) \\ 
MTM && 0.013 (0.408, 0.415) & 0.011 (0.264, 0.254) & -0.015 (0.875, 0.634) & 0.005 (0.349, 0.268)  & -0.027 (0.420, 0.303)\\

LIK & 0.6 & -0.337 (0.203, 0.211) &  -0.125 (0.168, 0.164) & -0.826 (0.234, 0.211) & -0.237 (0.147, 0.106) & 0.174 (0.228, 0.183) \\ 
MTM && 0.004 (0.418, 0.399) & 0.026 (0.259, 0.249) & -0.047 (0.860, 0.598) & 0.005 (0.360, 0.257)  & -0.002 (0.325, 0.216)\\

LIK & 0.8 & -0.334 (0.212, 0.212) &  -0.130 (0.173, 0.163) & -0.822 (0.241, 0.219) & -0.238 (0.144, 0.106) & 0.224 (0.172, 0.143) \\ 
MTM && -0.005 (0.413, 0.388) & 0.010 (0.261, 0.244) & -0.092 (0.849, 0.562) & -0.021 (0.320, 0.236)  & -0.014 (0.224, 0.125)\\

\end{tabular}
\end{center}
\label{tab:1}
\end{table}%
\end{landscape}

\begin{landscape}
\begin{table}[htp]
\caption{Bias (standard deviation s.d., average standard error s.e.) for the estimators of $\bar\eta, \bar\xi, \sigma^2_\eta, \sigma^2_\xi, \rho$ obtained from the approximate likelihood approach (LIK) and from the MTM approach (MTM), under increasing $\rho$ and different link function. High accuracy of the test. Sample size $n=10$. Prevalence of disease $\pi=0.35$.}
\begin{center}
\begin{tabular}{c c c c c c c}
Method & $\rho$ & $\bar\eta$ & $\bar\xi$& $\sigma^2_\eta$& $\sigma^2_\xi$&  $\rho$\\
&& Bias (s.d., s.e.) & Bias (s.d., s.e.) & Bias (s.d., s.e.) & Bias (s.d., s.e.)& Bias (s.d., s.e.) \\
\hline
{\it Logit link} &&&&&&\\
LIK & 0.2 & -0.432 (0.328, 0.367) &  -0.093 (0.240, 0.246) & -0.553 (0.460, 0.372) & -0.121 (0.252, 0.213) & 0.057 (0.338, 0.252) \\ 
MTM  && -0.017 (0.422, 0.426) & 0.013 (0.271, 0.271) & -0.144 (0.681, 0.557) & -0.012 (0.362, 0.285) & 0.016 (0.556, 0.365)\\
LIK & 0.6 & -0.369 (0.330, 0.354) &  -0.054 (0.249, 0.242) & -0.559 (0.487, 0.355) & -0.110 (0.285, 0.218) & 0.164 (0.270, 0.213) \\ 
MTM  && 0.024 (0.437, 0.413) & 0.010 (0.283, 0.271) & -0.144 (0.651, 0.550) & -0.025 (0.345, 0.288) & -0.002 (0.444, 0.291)\\

LIK & 0.8 & -0.356 (0.331, 0.355) &  -0.015 (0.235, 0.248) & -0.549 (0.490, 0.366) & -0.078 (0.290, 0.236) & 0.241 (0.245, 0.182) \\ 
MTM  && 0.039 (0.466, 0.416) & 0.027 (0.291, 0.281) & -0.041 (0.727, 0.602) & 0.020 (0.367, 0.325) & 0.021 (0.358, 0.201)\\
{\it Probit link} &&&&&&\\
LIK & 0.2 & -0.193 (0.276, 0.255) &  -0.041 (0.203, 0.190) & -0.563 (0.359, 0.253) & -0.152 (0.154, 0.122) & 0.041 (0.328, 0.243) \\ 
MTM  && -0.006 (0.396, 0.399) & 0.002 (0.260, 0.260) & -0.048 (0.728, 0.588) & -0.024 (0.322, 0.267) & -0.018 (0.448, 0.306)\\

LIK & 0.6 & -0.185 (0.281, 0.253) &  -0.036 (0.194, 0.187) & -0.569 (0.362, 0.254) & -0.156 (0.155, 0.123) & 0.139 (0.245, 0.193) \\ 
MTM  && 0.001 (0.399, 0.393) & 0.009 (0.256, 0.260) & -0.078 (0.737, 0.547) & -0.030 (0.327, 0.258) & -0.016 (0.327, 0.219)\\

LIK & 0.8 & -0.195 (0.290, 0.250) &  -0.032 (0.201, 0.187) & -0.590 (0.347, 0.240) & -0.158 (0.159, 0.124) & 0.180 (0.191, 0.150) \\ 
MTM  && -0.027 (0.391, 0.372) & 0.024 (0.255, 0.259) & -0.162 (0.667, 0.493) & -0.032 (0.318, 0.242) & -0.029 (0.230, 0.126)\\

{\it Cloglog link} &&&&&&\\
LIK & 0.2 & -0.298 (0.231, 0.223) &  -0.152 (0.169, 0.163) & -0.726 (0.280, 0.222) & -0.246 (0.144, 0.106) & 0.047 (0.321, 0.235) \\ 
MTM  && 0.014 (0.403, 0.400) & 0.021 (0.269, 0.265) & -0.013 (0.819, 0.610) & 0.026 (0.378, 0.284) & -0.015 (0.412, 0.304)\\
LIK & 0.6 & -0.291 (0.234, 0.219) &  -0.141 (0.167, 0.161) & -0.733 (0.278, 0.218) & -0.250 (0.140, 0.107) & 0.147 (0.241, 0.181) \\ 
MTM  && -0.009 (0.384, 0.387) & 0.017 (0.262, 0.254) & -0.082 (0.758, 0.565) & -0.009 (0.333, 0.258) & -0.006 (0.316, 0.217)\\
LIK & 0.8 & -0.278 (0.228, 0.219) &  -0.148 (0.165, 0.162) & -0.725 (0.283, 0.225) & -0.245 (0.139, 0.104) & 0.198 (0.168, 0.137) \\ 
MTM  && 0.006 (0.399, 0.379) & 0.005 (0.263, 0.253) & -0.092 (0.787, 0.515) & -0.014 (0.330, 0.237) & -0.013 (0.205, 0.125)\\
\end{tabular}
\end{center}
\label{tab:2}
\end{table}%
\end{landscape}

\section{Application}
Delirium is an acute confusional state with varying disturbances of cognition, memory, attention, behaviour, and orientation. It is often observed in early stages of the hospitalization for acute and chronic diseases, especially in the elderly (Lipowski, 1987; Rai et al., 2014). Since delirium has been associated with unfavorable outcomes, early recognition and prompt treatment is crucial to decrease the risk of morbidity and/or mortality.
Shi et al. (2013) perform a meta-analysis of diagnostic accuracy of the confusion assessment method, which is one of the most widely used delirium screening tool (Inouye et al., 1990) used by non-psychiatrically trained clinicians (nurses, general practitioners, ...) to identify and recognize delirium quickly. The confusion assessment method can be applied to verbal and nonverbal (e.g., mechanically ventilated) patients. It is considered as an alternative to the golden standard diagnostic criterion, namely, the Diagnostic and Statistical Manual of Mental Disorders IV, which cannot be easily applied to daily bedside practice. Table~\ref{Table:example} reports the the data for 20 studies included in the meta-analysis.

\begin{table}[ht]
\caption{Data for the confusion assessment method example (Shi et al., 2013). TP=true positives, FP=false positives, FN=false negatives, TN=true negatives.}
\centering
\begin{tabular}{rrrrrr}
  \hline
 & Study & TP & FP & TN & FN \\ 
  \hline
1 & 1 & 21 & 4 & 43 & 3 \\ 
  2 & 2 & 35 & 2 & 104 & 40 \\ 
  3 & 3 & 77 & 0 & 16 & 3 \\ 
  4 & 4 & 16 & 3 & 80 & 1 \\ 
  5 & 5 & 27 & 0 & 93 & 3 \\ 
  6 & 6 & 22 & 0 & 19 & 3 \\ 
  7 & 7 & 33 & 3 & 70 & 13 \\ 
  8 & 8 & 26 & 8 & 41 & 6 \\ 
  9 & 9 & 19 & 0 & 75 & 6 \\ 
  10 & 10 & 22 & 2 & 76 & 2 \\ 
  11 & 11 & 137 & 11 & 36 & 42 \\ 
  12 & 12 & 225 & 12 & 706 & 60 \\ 
  13 & 13 & 14 & 3 & 344 & 62 \\ 
  14 & 14 & 9 & 2 & 131 & 12 \\ 
  15 & 15 & 15 & 0 & 71 & 2 \\ 
  16 & 16 & 15 & 0 & 37 & 2 \\ 
  17 & 17 & 39 & 13 & 64 & 16 \\ 
  18 & 18 & 15 & 0 & 58 & 6 \\ 
  19 & 19 & 56 & 6 & 59 & 5 \\ 
  20 & 20 & 34 & 0 & 29 & 3 \\ 
   \hline
\end{tabular}
\label{Table:example}
\end{table}

The hierarchical MTM and the standard approximate bivariate model are applied to the data under a logit specification of the relationship between sensitivity and specificity of the diagnostic tool. The choice is motivated by the reduced value of the Akaike Information Criterion associate to the logit specification if compared to the probit and the cloglog alternatives.
Results are reported in Table~\ref{Table:results}. 
\begin{table}[ht]
\caption{Results for the confusion assessment method example (Shi et al., 2013). Estimates and standard error (in parentheses) of the parameters of the MTM and the approximate bivariate random-effects model and estimate (and standard error) of sensitivity and specificity of the diagnostic tool.}
\centering
\begin{tabular}{cccccccc}
  \hline
  Model & $\bar\eta$ & $\bar\xi$ &  $\sigma^2_\eta$ & $\sigma^2_\xi$ & $\rho$ & SE & SP \\ 
 \hline
 MTM & 1.44 (0.25) &  3.67 (0.30)   &  1.17 (0.49)   & 1.45 (0.35)  & -0.10 (0.20)& 0.81 (0.04)& 0.98 (0.01)\\
 Approximate &&&&&&&\\
 likelihood &1.38 (0.23)   & 3.25 (0.27)  &  1.06 (0.47)  &  1.06 (0.27) &  -0.21 (0.18)& 0.80 (0.04) & 0.96 (0.01)\\
  \hline
\end{tabular}
\label{Table:results}
\end{table}

\begin{figure}[hbtp]
\begin{center}
\includegraphics[width=4.5in]{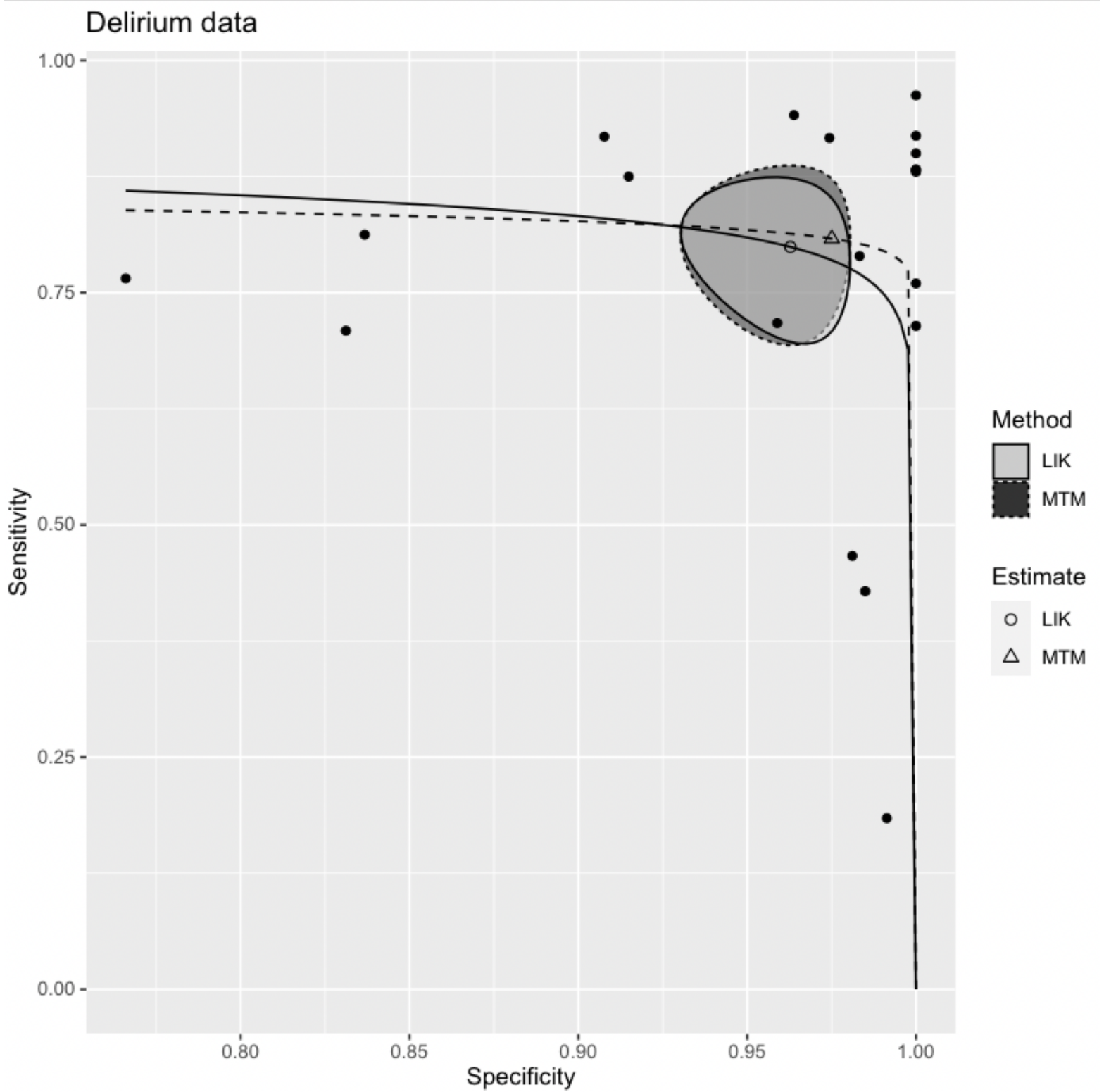}
\caption{Summary ROC curves, estimated sensitivity and specificity, and associated 95\% confidence regions from approximate likelihood and MTM approach
 for the confusion assessment method example. (Shi et al., 2013).} 
\label{Fig:example}
\end{center}
\end{figure}

No convergence problem has been experienced for both the approaches. The hierarchical MTM provides larger estimates of the variance components and a smaller value of the correlation between sensitivity and specificity, at the price of a slightly larger standard error as a consequence of the increased complexity of the model if compared to the approximate likelihood approach. Globally, sensitivity and specificity of the competing methods are close, with specificity larger than sensitivity. In particular, the MTM approach provides an estimate of sensitivity and specificity equal to 0.81 (standard error 0.04) and 0.98 (standard error 0.01), respectively. The likelihood approach for the bivariate model provides an estimate of sensitivity and specificity equal to 0.80 (standard error 0.04) and to 0.96 (standard error 0.01), respectively.
Figure~\ref{Fig:example} reports the summary ROC curves from each method (Arends et al., 2008) the estimated sensitivity and specificity, together with the associated 95\% confidence region. Differences among the summary ROC curves and among the 95\% confidence regions are slight and reflect the slight differences in the estimated sensitivity and specificity from the approaches. 

\section{Conclusions} 
This paper explores the use of multinomial tree models, an instrument often adopted in psychological research to investigate cognitive processes, to carry out meta-analysis of diagnostic accuracy studies. The focus is on the extension of the fixed-effects model developed in Botella et al. (2013) to a hierarchical structure accounting for heterogeneity among studies included in the meta-analysis. In this way, the model meets the random-effects formulation commonly adopted in meta-analysis and allows to properly distinguish within-study and between-study heterogeneity. Inference is then performed from a frequentist perspective, in contrasts to the standard latent-trait approach usually based on Bayesian solutions (Klauer, 2010). %Whichever the link used to translate the test accuracy measure SE and SP to the real line, the resulting likelihood function is not in closed-form, thus requiring numerical integration for inference to carry out. 
Interestingly, it is shown that the associated likelihood function for the parameters expressing the accuracy measures resembles the likelihood function obtained under a classical bivariate random-effects approach to meta-analysis of diagnostic tests under an exact specification of the distribution for the true positives and the true negatives classified by the test under study (Arends et al., 2008). The added value of the MTM specification is the possibility to express and estimate the study-specific prevalences of the disease, by exploiting the parameters' separability of the complete likelihood function. 
Simulation studies under a variety of scenarios show that the MTM likelihood-based approach is preferable to the classical likelihood solution constructed on the approximate normal distribution for the sensitivity and specificity of the test in terms of accuracy of the inferential results, especially in case of small number of studies included in the meta-analysis. The only disadvantage is represented by the need of numerical integration, which can be easily solved via quadrature methods.

The multinomial processing tree model has been proposed and adapted in case the accuracy of the test is compared to that of a perfect reference standard. When the reference test is imperfect, an interesting extension of the multinomial tree structure would include different sensitivities and specificities for the test under study and the reference, with an expected complication of the model structure (Botella et al., 2013). How the hierarchical model and the associated likelihood change in this case and compare with the classical bivariate approximate solution represents an interesting future direction of the present work.

\section{Software}
\label{sec5}

Software in the form of \texttt R code %, together with a sample
%input data set and complete documentation is available on
%request from the corresponding author (eaheron@tcd.ie).
is available at  https://github.com/annamariaguolo/MTM-meta-analysis.

%\section{Supplementary Material}
%\label{sec6}

%Supplementary material is available online at
%\href{http://biostatistics.oxfordjournals.org}%
%{http://biostatistics.oxfordjournals.org}.

\section*{Acknowledgments}
The Author is grateful to Dr. Jessica Battagello for helpful discussion.

%{\it Conflict of Interest}: None declared.

%\bibliographystyle{biorefs}
%\bibliography{refs}
%\bibliographystyle{apalike}
%\bibliography{Guolo_bibliography}
%\bibliography{Guolo_Main_Paper.bbl}

\begin{thebibliography}{10}

\bibitem{arends2008}
Arends, L., Hamza, T., van Houwelingen, H., Heijenbrok-Kal, M., Hunink, M., and Stijnen, T. (2008). Bivariate random effects meta-analysis of roc curves. {\it Med Decis Making}, {\bf 28}: 621--638.
\bibitem{batchelder1999}
Batchelder, W. and Riefer, D. (1999). Theoretical and empirical review of multinomial process tree modeling. {\it Psychon B Rev}, {\bf 6}: 57--86.
\bibitem{botella2013}
Botella, J., Huang, H., and Suero, M. (2013). Multinomial tree models for assessing the status of the reference in studies of the accuracy of tools for binary classification. {\it Front Psychol}, {\bf 4}: 694.
\bibitem{chen2017}
Chen, Y., Liu, Y., Ning, J., Nie, L., Zhu, H., and Chu, H. (2017). A composite likelihood method for bivariate meta-analysis in diagnostic systematic reviews. {\it Stat Methods Med Res}, {\bf 26}: 914--930.
\bibitem{chu2006}
Chu, H. and Cole, S. (2006). Bivariate meta-analysis of sensitivity and specificity with sparse data: a generalized linear mixed model approach. {\it J Clin Epidemiol}, {\bf 59}: 1331--1333.
\bibitem{erdfelder2009}
Erdfelder, E., Auer, T., Hilbig, B., Aßfalg, A., Moshagen, M., and Nadarevic, L. (2009). Multinomial processing tree models: A review of the literature. {\it J Psychol}, {\bf 217}: 108--124.
\bibitem{guolo2017}
Guolo, A. (2017). A double simex approach for bivariate random-effects meta-analysis of diagnostic accuracy studies. {\it BMC Med Res Methodol}, {\bf 17}: 6.
\bibitem{hamza2008}
Hamza, T., van Houwelingen, H., and Stijnen, T. (2008). The binomial distribution of meta-analysis was preferred to model within-study variability. {\it J Clin Epidemiol}, {\bf 61}: 41--51.
\bibitem{heck2018}
Heck, D., Arnold, N., and Arnold, D. (2018). Treebugs: An r package for hierarchical multinomial- processing-tree modeling. {\it Behav Res Methods}, {\bf 50}: 264--284.
\bibitem{honest2002}
Honest, H. and Khan, K. (2002). Reporting of measures of accuracy in systematic reviews of diagnostic literature. {\it BMC Health Serv Res}, {\bf 2}: 4.
\bibitem{inouye1990}
Inouye, S., van Dyck, C., Alessi, C., Balkin, S., Siegal, A., and Horwitz, R. (1990). Clarifying confusion: the confusion assessment method. a new method for detection of delirium. {\it Ann Intern Med}, {\bf 113}: 941--948.
\bibitem{jackson2011}
Jackson, D., Riley, R., and White, I. (2011). Multivariate meta-analysis: Potential and promise. {\it Stat Med}, {\bf 30}: 2481--2498.
\bibitem{jobst2020}
Jobst, L., Heck, D., and Moshagen, M. (2020). A comparison of correlation and regression approaches for multinomial processing tree models. {\it J Math Psychol}, {\bf 98}: 102400.
\bibitem{kauermann2001}
Kauermann, G. and Carroll, R. (2001). A note on the efficiency of sandwich covariance matrix estimation. {\it J Am Stat Assoc}, {\bf 96}: 1387--1396.
\bibitem{klauer2006}
Klauer, K. (2006). Hierarchical multinomial processing tree models: a latent-class approach. {\it Psychometrika}, {\bf 71}: 7--31.
\bibitem{klauer2010}
Klauer, K. (2010). Hierarchical multinomial processing tree models: a latent-trait approach. {\it Psychometrika}, {\bf 75}: 70--98.
\bibitem{lipowski1987}
Lipowski, Z. (1987). Delirium (acute confusional states). {\it J Amer Medical Assoc}, 258:1789?1792. 
\bibitem{littenberg1993}
Littenberg, B. and Moses, L. (1993). Estimating diagnostic-accuracy from multiple conflicting reports: a new meta-analytic method. {\it Med Decis Making}, {\bf 13}: 313--321.
\bibitem{moses1993}
Moses, L., Shapiro, D., and Littenberg, B. (1993). Combining independent studies of a diagnostic test into a summary roc curve: data-analytic approaches and some additional consideration. {\it Stat Med}, {\bf 12}: 1293--1316.
\bibitem{nelder1965}
Nelder, J. and Mead, R. (1965). A simplex algorithm for function minimization. {\it Scand J Stat}, {\bf 7}: 308--313.
\bibitem{r2022}
R Core Team (2022). R: A language and environment for statistical computing. {\it R Foundation for Statistical Computing}, Vienna, Austria, https://www.R-project.org/.
\bibitem{rai2014}
Rai, D., Garg, R., Malhotra, H., Verma, R., Jain, A., Tiwari, S., and Signh, M. (2014). Acute confusional state/delirium: An etiological and prognostic evaluation. {\it Ann Indian Acad Neurol}, {\bf 17}: 30--34.
\bibitem{reitsma2005}
Reitsma, J., Glas, A., Rutjes, A., Scholten, RJ, B., PM, and Zwinderman, A. (2005). Bivariate analysis of sensitivity and specificity produces informative summary measures in diagnostic reviews. {\it J Clin Epidemiol}, {\bf 58}: 982--990.
\bibitem{riefer1988}
Riefer, D. and Batchelder, W. (1998). Multinomial modeling and the measurement of cognitive processes. {\it Psychol Rev}, {\bf 95}: 318--339.
\bibitem{rutter2001}
Rutter, C. and Gatsonis, C. (2001). A hierarchical regression approach to meta- analysis of diagnostic test accuracy evaluations. {\it Stat Med}, {\bf 20}: 2865--2884.
\bibitem{shi2013}
Shi, Q., Warren, L., Saposnik, G., and MacDermid, J. (2013). Confusion assessment method: a systematic review and meta-analysis of diagnostic accuracy. {\it Neuropsych Dis Treat}, {\bf 9}: 1359--1370.
\bibitem{stahl2007}
Stahl, C. and Klauer, K. (2007). Hmmtree: A computer program for latent-class hierarchical multinomial processing tree models. {\it Behav Res Methods}, {\bf 39}: 267--273.
\bibitem{takwoingi2017}
Takwoingi, Y., Guo, B., Riley, R., and Deeks, J. (2017). Performance of methods for meta-analysis of diagnostic test accuracy with few studies or sparse data. {\it Stat Methods Med Res}, {\bf 26}: 1896--1911.

\end{thebibliography}

\newpage
\clearpage

\pagestyle{plain}

\begin{center}
{\bf \large Supplementary Material for \\
Hierarchical multinomial processing tree models for meta-analysis of diagnostic accuracy studies} 
\\[2ex]
Annamaria Guolo\footnote{Department of Statistical Sciences, Via Cesare Battisti 241/243, Padova, Italy; I-35128; annamaria.guolo@unipd.it}
\\
Department of Statistical Sciences, University of Padova
\end{center}

\renewcommand{\thetable}{S\arabic{table}}
\renewcommand{\thefigure}{S\arabic{figure}}

\section*{Additional simulation results}
This section reports a portion of the results of the simulation study carried out to compare the performance of the competing approaches, as described in Section 4 of the main manuscript.  

% high accuracy, n 25 , pi 20
\begin{landscape}
\begin{table}[htp]
\caption{Bias (standard deviation s.d., average standard error s.e.) for the estimators of $\bar\eta, \bar\xi, \sigma^2_\eta, \sigma^2_\xi, \rho$ obtained from the approximate likelihood approach (LIK) and from the MTM approach (MTM), under increasing $\rho$ and different link function. High accuracy of the test. Sample size $n=25$. Prevalence of disease $\pi=0.20$.}
\begin{center}
\begin{tabular}{c c c c c c c}
Method & $\rho$ & $\bar\eta$ & $\bar\xi$& $\sigma^2_\eta$& $\sigma^2_\xi$&  $\rho$\\
&& Bias (s.d., s.e.) & Bias (s.d., s.e.) & Bias (s.d., s.e.) & Bias (s.d., s.e.)& Bias (s.d., s.e.) \\
\hline
{\it Logit link} &&&&&&\\
LIK & 0.2 & -0.579 (0.204, 0.219) &  -0.078 (0.157, 0.149) & -0.716 (0.299, 0.240) & -0.100 (0.159, 0.145) & 0.079 (0.214, 0.188) \\ 
MTM &  & -0.035 (0.281, 0.289) &  0.000 (0.170, 0.160) & -0.160 (0.486, 0.411) & -0.014 (0.188, 0.175) & -0.014 (0.350, 0.277) \\ 
LIK & 0.6 & -0.543 (0.199, 0.215) &  -0.028 (0.152, 0.150) & -0.709 (0.295, 0.240) & -0.075 (0.178, 0.157) & 0.196 (0.182, 0.158) \\ 
MTM &  & -0.037 (0.289, 0.287) &  0.004 (0.169, 0.162) & -0.165 (0.488, 0.420) & -0.007 (0.209, 0.192) & -0.039 (0.278, 0.213) \\ 
LIK & 0.8 & -0.515 (0.201, 0.213) &  -0.005 (0.150, 0.153) & -0.700 (0.295, 0.240) & -0.039 (0.188, 0.170) & 0.263 (0.151, 0.134) \\ 
MTM &  & -0.013 (0.303, 0.285) &  0.019 (0.180, 0.171) & -0.100 (0.465, 0.452) & 0.034 (0.235, 0.218) & -0.039 (0.193, 0.149) \\ 

{\it Probit link} &&&&&&\\
LIK & 0.2 & -0.248 (0.173, 0.160) &  -0.024 (0.128, 0.127) & -0.642 (0.237, 0.190) & -0.110 (0.105, 0.096) & 0.051 (0.192, 0.175) \\ 
MTM &  & 0.009 (0.275, 0.271) &  0.006 (0.158, 0.158) & 0.001 (0.529, 0.466) & -0.011 (0.197, 0.176) & -0.004 (0.255, 0.228) \\ 
LIK & 0.6 & -0.247 (0.166, 0.158) &  -0.030 (0.127, 0.128) & -0.666 (0.216, 0.182) & -0.108 (0.107, 0.096) & 0.152 (0.144, 0.134) \\ 
MTM &  & 0.002 (0.267, 0.264) &  0.004 (0.155, 0.158) & -0.047 (0.517, 0.422) & -0.009 (0.192, 0.173) & -0.012 (0.194, 0.162) \\ 
LIK & 0.8 & -0.262 (0.173, 0.160) &  -0.027 (0.137, 0.128) & -0.650 (0.223, 0.185) & -0.110 (0.103, 0.094) & 0.205 (0.106, 0.102) \\ 
MTM &  & -0.020 (0.272, 0.254) &  0.011 (0.167, 0.158) & -0.060 (0.508, 0.429) & -0.006 (0.186, 0.176) & -0.026 (0.131, 0.096) \\ 

{\it Cloglog link} &&&&&&\\
LIK & 0.2 & -0.334 (0.139, 0.137) &  -0.134 (0.112, 0.108) & -0.821 (0.158, 0.146) & -0.221 (0.097, 0.082) & 0.041 (0.184, 0.165) \\ 
MTM &  & 0.013 (0.263, 0.269) &  0.008 (0.162, 0.159) & 0.025 (0.572, 0.511) & 0.011 (0.221, 0.184) & -0.016 (0.241, 0.218) \\ 
LIK & 0.6 & -0.324 (0.138, 0.134) &  -0.131 (0.108, 0.107) & -0.827 (0.153, 0.144) & -0.222 (0.095, 0.080) & 0.180 (0.137, 0.121) \\ 
MTM &  & 0.012 (0.273, 0.266) &  0.006 (0.153, 0.159) & -0.005 (0.565, 0.490) & 0.006 (0.208, 0.177) & 0.009 (0.188, 0.158) \\
 LIK & 0.8 & -0.320 (0.135, 0.133) &  -0.129 (0.107, 0.107) & -0.831 (0.149, 0.142) & -0.222 (0.090, 0.081) & 0.232 (0.099, 0.090) \\ 
MTM &  & -0.005 (0.259, 0.252) &  0.011 (0.159, 0.156) & -0.057 (0.575, 0.455) & -0.001 (0.197, 0.174) & -0.016 (0.123, 0.092) \\
\end{tabular}
\end{center}
%\label{tab:1}
\end{table}%

% high accuracy, n 25 , pi 35
\begin{table}[htp]
\caption{Bias (standard deviation s.d., average standard error s.e.) for the estimators of $\bar\eta, \bar\xi, \sigma^2_\eta, \sigma^2_\xi, \rho$ obtained from the approximate likelihood approach (LIK) and from the MTM approach (MTM), under increasing $\rho$ and different link function. High accuracy of the test. Sample size $n=25$. Prevalence of disease $\pi=0.35$.}
\begin{center}
\begin{tabular}{c c c c c c c}
Method & $\rho$ & $\bar\eta$ & $\bar\xi$& $\sigma^2_\eta$& $\sigma^2_\xi$&  $\rho$\\
&& Bias (s.d., s.e.) & Bias (s.d., s.e.) & Bias (s.d., s.e.) & Bias (s.d., s.e.)& Bias (s.d., s.e.) \\
\hline
{\it Logit link} &&&&&&\\
LIK & 0.2 & -0.409 (0.215, 0.213) &  -0.093 (0.154, 0.149) & -0.599 (0.311, 0.254) & -0.141 (0.154, 0.139) & 0.076 (0.206, 0.182) \\ 
MTM && -0.025 (0.268, 0.261) & 0.003 (0.169, 0.162) & -0.144 (0.427, 0.377) & -0.034 (0.189, 0.177) & -0.012 (0.306, 0.260) \\
LIK & 0.6 & -0.363 (0.212, 0.214) &  -0.055 (0.156, 0.153) & -0.579 (0.308, 0.255) & -0.100 (0.177, 0.153) & 0.196 (0.175, 0.156) \\ 
MTM && -0.011 (0.273, 0.263) & 0.002 (0.177, 0.167) & -0.108 (0.432, 0.383) & 0.001 (0.218, 0.191) & -0.019 (0.243, 0.201) \\
LIK & 0.8 & -0.339 (0.202, 0.211) &  -0.029 (0.148, 0.151) & -0.557 (0.328, 0.262) & -0.085 (0.181, 0.158) & 0.249 (0.144, 0.128) \\ 
MTM && 0.009 (0.271, 0.265) & 0.011 (0.176, 0.168) & -0.057 (0.461, 0.429) & 0.014 (0.228, 0.209) & -0.032 (0.174, 0.139) \\

{\it Probit link} &&&&&&\\
LIK & 0.2 & -0.188 (0.168, 0.168) &  -0.034 (0.129, 0.124) & -0.529 (0.224, 0.195) & -0.129 (0.103, 0.093) & 0.040 (0.187, 0.174) \\ 
MTM && -0.002 (0.245, 0.253) & 0.005 (0.168, 0.162) & -0.006 (0.478, 0.431) & -0.007 (0.212, 0.180) & -0.010 (0.242, 0.221) \\
LIK & 0.6 & -0.184 (0.166, 0.167) &  -0.036 (0.127, 0.126) & -0.543 (0.224, 0.194) & -0.123 (0.103, 0.095) & 0.137 (0.140, 0.132) \\ 
MTM && -0.007 (0.243, 0.250) & 0.006 (0.160, 0.163) & -0.024 (0.470, 0.426) & 0.001 (0.198, 0.181) & -0.010 (0.182, 0.157) \\
LIK & 0.8 & -0.183 (0.176, 0.166) &  -0.036 (0.129, 0.125) & -0.554 (0.233, 0.189) & -0.132 (0.104, 0.093) & 0.189 (0.109, 0.099) \\ 
MTM && -0.001 (0.248, 0.241) & 0.006 (0.164, 0.159) & -0.070 (0.489, 0.397) & -0.013 (0.194, 0.175) & -0.017 (0.129, 0.097) \\

{\it Cloglog link} &&&&&&\\

LIK & 0.2 & -0.286 (0.146, 0.146) &  -0.149 (0.109, 0.106) & -0.707 (0.172, 0.163) & -0.237 (0.092, 0.080) & 0.048 (0.187, 0.167) \\ 
MTM && 0.012 (0.252, 0.256) & 0.001 (0.163, 0.160) & 0.016 (0.519, 0.467) & 0.000 (0.209, 0.184) & 0.001 (0.235, 0.215) \\
LIK & 0.6 & -0.282 (0.145, 0.144) &  -0.146 (0.106, 0.105) & -0.710 (0.174, 0.161) & -0.236 (0.090, 0.080) & 0.156 (0.130, 0.120) \\ 
MTM && 0.006 (0.255, 0.253) & 0.002 (0.163, 0.160) & 0.003 (0.516, 0.455) & 0.000 (0.209, 0.183) & -0.001 (0.173, 0.153) \\
LIK & 0.8 & -0.272 (0.142, 0.143) &  -0.141 (0.107, 0.105) & -0.706 (0.182, 0.164) & -0.235 (0.095, 0.080) & 0.218 (0.093, 0.088) \\ 
MTM && 0.006 (0.248, 0.247) & 0.005 (0.166, 0.157) & -0.025 (0.525, 0.430) & -0.002 (0.212, 0.174) & -0.005 (0.113, 0.093) \\

\end{tabular}
\end{center}
%\label{tab:1}
\end{table}%

% low accuracy, n 10 , pi 20

\begin{table}[htp]
\caption{Bias (standard deviation s.d., average standard error s.e.) for the estimators of $\bar\eta, \bar\xi, \sigma^2_\eta, \sigma^2_\xi, \rho$ obtained from the approximate likelihood approach (LIK) and from the MTM approach (MTM), under increasing $\rho$ and different link function. Low accuracy of the test. Sample size $n=10$. Prevalence of disease $\pi=0.20$.}
\begin{center}
\begin{tabular}{c c c c c c c}
Method & $\rho$ & $\bar\eta$ & $\bar\xi$& $\sigma^2_\eta$& $\sigma^2_\xi$&  $\rho$\\
&& Bias (s.d., s.e.) & Bias (s.d., s.e.) & Bias (s.d., s.e.) & Bias (s.d., s.e.)& Bias (s.d., s.e.) \\
\hline
{\it Logit link} &&&&&&\\
LIK & 0.2 & -0.202 (0.339, 0.334) &  -0.030 (0.231, 0.223) & -0.503 ( 0.466, 0.410) & -0.094 (0.241, 0.192) & 0.041 (0.322, 0.245) \\ 
MTM && 0.000 (0.414, 0.390) &  0.003 (0.245, 0.235) & -0.101 ( 0.745, 0.609) & -0.038 (0.284, 0.228) & -0.013 (0.461, 0.322) \\ 
LIK & 0.6 & -0.193 (0.339, 0.338) &  -0.001 (0.222, 0.223) & -0.486 (0.487, 0.425) & -0.092 (0.236, 0.198) & 0.144 (0.273, 0.199) \\ 
MTM && 0.001 (0.423, 0.393) & 0.011 (0.247, 0.236) & -0.037 (0.766, 0.629) & -0.002 (0.288, 0.242) & -0.001 (0.366, 0.231) \\
LIK & 0.8 & -0.145 (0.325, 0.338) &  -0.011 (0.230, 0.222) & -0.449 (0.516, 0.435) & -0.078 (0.265, 0.201) & 0.174 (0.206, 0.160) \\ 
MTM && 0.048 (0.411, 0.398) & -0.012 (0.260, 0.245) & 0.056 (0.803, 0.665) & 0.007 (0.341, 0.271) & -0.002 (0.259, 0.160) \\

{\it Probit link} &&&&&&\\
LIK & 0.2 & -0.056 (0.321, 0.298) &  0.013 (0.220, 0.207) & -0.335 (0.405, 0.308) & -0.069 (0.177, 0.154) & 0.032 (0.329, 0.244) \\ 
MTM && 0.013 (0.379, 0.373) & 0.010 (0.243, 0.236) & -0.046 (0.771, 0.601) & -0.032 (0.273, 0.238) & 0.000 (0.395, 0.296) \\
LIK & 0.6 & -0.062 (0.330, 0.302) &  0.008 (0.228, 0.209) & -0.321 (0.394, 0.315) & -0.059 (0.202, 0.157) & 0.099 (0.258, 0.189) \\ 
MTM && 0.016 (0.387, 0.375) & 0.011 (0.249, 0.240) & -0.020 (0.768, 0.604) & -0.012 (0.308, 0.239) & 0.014 (0.307, 0.214) \\
LIK & 0.8 & -0.074 (0.323, 0.302) &  0.017 (0.226, 0.206) & -0.317 (0.403, 0.306) & -0.075 (0.180, 0.150) & 0.117 (0.191, 0.136) \\ 
MTM && -0.006 (0.370, 0.369) & 0.016 (0.246, 0.230) & -0.061 (0.701, 0.567) & -0.046 (0.262, 0.223) & 0.006 (0.208, 0.137) \\

{\it Cloglog link} &&&&&&\\

LIK & 0.2 & -0.122 (0.281, 0.263) &  -0.070 (0.182, 0.178) & -0.620 (0.311, 0.275) & -0.195 (0.161, 0.119) & 0.039 (0.319, 0.241) \\ 
MTM && 0.012 (0.400, 0.386) & 0.019 (0.237, 0.239) & 0.046 (0.867, 0.654) & -0.015 (0.314, 0.249) & 0.008 (0.390, 0.293) \\
LIK & 0.6 & -0.103 (0.276, 0.256) &  -0.085 (0.189, 0.179) & -0.650 (0.286, 0.259) & -0.195 (0.149, 0.120) & 0.117 (0.231, 0.190) \\ 
MTM && 0.011 (0.392, 0.371) & 0.008 (0.249, 0.237) & -0.045 (0.767, 0.601) & -0.021 (0.276, 0.240) & 0.019 (0.281, 0.217) \\
LIK & 0.8 & -0.092 (0.264, 0.257) &  -0.086 (0.184, 0.180) & -0.653 (0.281, 0.259) & -0.195 (0.158, 0.118) & 0.141 (0.174, 0.137) \\ 
MTM && 0.013 (0.381, 0.370) & 0.008 (0.240, 0.236) & -0.044 (0.779, 0.565) & -0.013 (0.301, 0.233) & 0.006 (0.202, 0.131) \\

\end{tabular}
\end{center}

\end{table}%

% low accuracy, n 10 , pi 35

\begin{table}[htp]
\caption{Bias (standard deviation s.d., average standard error s.e.) for the estimators of $\bar\eta, \bar\xi, \sigma^2_\eta, \sigma^2_\xi, \rho$ obtained from the approximate likelihood approach (LIK) and from the MTM approach (MTM), under increasing $\rho$ and different link function. Low accuracy of the test. Sample size $n=10$. Prevalence of disease $\pi=0.35$.}
\begin{center}
\begin{tabular}{c c c c c c c}
Method & $\rho$ & $\bar\eta$ & $\bar\xi$& $\sigma^2_\eta$& $\sigma^2_\xi$&  $\rho$\\
&& Bias (s.d., s.e.) & Bias (s.d., s.e.) & Bias (s.d., s.e.) & Bias (s.d., s.e.)& Bias (s.d., s.e.) \\
\hline
{\it Logit link} &&&&&&\\
LIK & 0.2 & -0.120 (0.332, 0.327) &  -0.041 (0.234, 0.227) & -0.387 ( 0.476, 0.414) & -0.110 (0.228, 0.200) & 0.066 (0.312, 0.240) \\ 
MTM && 0.005 (0.378, 0.365) &  0.009 (0.252, 0.240) & -0.102 ( 0.648, 0.543) & -0.042 (0.274, 0.238) & 0.026 (0.427, 0.306)\\
LIK & 0.6 & -0.100 (0.338, 0.328) &  -0.029 (0.233, 0.223) & -0.359 (0.548, 0.411) & -0.116 (0.238, 0.190) & 0.136 (0.258, 0.198) \\ 
MTM && 0.012 (0.394, 0.370) & 0.000 (0.256, 0.240) & -0.048 (0.719, 0.570) & -0.045 (0.288, 0.238) & -0.002 (0.341, 0.222) \\
LIK & 0.8 & -0.083 (0.325, 0.327) &  -0.017 (0.237, 0.222) & -0.364 (0.508, 0.415) & -0.113 (0.235, 0.192) & 0.155 (0.191, 0.154) \\ 
MTM && 0.020 (0.389, 0.370) & 0.006 (0.273, 0.241) & 0.017 (0.740, 0.598) & -0.003 (0.316, 0.266) & -0.019 (0.219, 0.143) \\

{\it Probit link} &&&&&&\\
LIK & 0.2 & -0.048 (0.328, 0.304) &  0.007 (0.220, 0.207) & -0.268 (0.406, 0.322) & -0.067 (0.202, 0.154) & 0.029 (0.325, 0.244) \\ 
MTM && -0.007 (0.359, 0.356) & 0.007 (0.243, 0.242) & -0.079 (0.664, 0.547) & -0.011 (0.320, 0.252) & 0.005 (0.381, 0.283) \\
LIK & 0.6 & -0.032 (0.335, 0.306) &  0.004 (0.217, 0.204) & -0.257 (0.432, 0.329) & -0.081 (0.185, 0.148) & 0.094 (0.254, 0.189) \\ 
MTM && 0.016 (0.370, 0.356) & 0.002 (0.241, 0.237) & -0.054 (0.706, 0.549) & -0.035 (0.286, 0.237) & 0.023 (0.284, 0.217) \\
LIK & 0.8 & -0.054 (0.329, 0.307) &  -0.005 (0.238, 0.204) & -0.246 (0.439, 0.324) & -0.088 (0.165, 0.144) & 0.104 (0.187, 0.131) \\ 
MTM && 0.003 (0.382, 0.354) & -0.008 (0.267, 0.237) & -0.045 (0.705, 0.531) & -0.037 (0.288, 0.235) & 0.002 (0.215, 0.125) \\

{\it Cloglog link} &&&&&&\\

LIK & 0.2 & -0.092 (0.287, 0.269) &  -0.087 (0.189, 0.177) & -0.530 (0.318, 0.280) & -0.205 (0.149, 0.118) & 0.041 (0.325, 0.241) \\ 
MTM && 0.032 (0.394, 0.368) & 0.012 (0.249, 0.242) & -0.019 (0.757, 0.588) & -0.017 (0.298, 0.254) & 0.009 (0.380, 0.289) \\
LIK & 0.6 & -0.105 (0.279, 0.269) &  -0.079 (0.181, 0.177) & -0.520 (0.325, 0.281) & -0.202 (0.158, 0.120) & 0.094 (0.228, 0.190) \\ 
MTM && 0.000 (0.379, 0.363) & 0.021 (0.247, 0.241) & -0.033 (0.724, 0.576) & -0.002 (0.342, 0.251) & 0.001 (0.272, 0.204) \\
LIK & 0.8 & -0.070 (0.276, 0.265) &  -0.095 (0.181, 0.178) & -0.540 (0.309, 0.279) & -0.199 (0.156, 0.121) & 0.124 (0.168, 0.132) \\ 
MTM && 0.030 (0.376, 0.356) & 0.000 (0.242, 0.236) & -0.061 (0.713, 0.538) & -0.014 (0.315, 0.236) & 0.011 (0.188, 0.126) \\

\end{tabular}
\end{center}

\end{table}%

% low accuracy, n 25 , pi 20

\begin{table}[htp]
\caption{Bias (standard deviation s.d., average standard error s.e.) for the estimators of $\bar\eta, \bar\xi, \sigma^2_\eta, \sigma^2_\xi, \rho$ obtained from the approximate likelihood approach (LIK) and from the MTM approach (MTM), under increasing $\rho$ and different link function. Low accuracy of the test. Sample size $n=25$. Prevalence of disease $\pi=0.20$.}
\begin{center}
\begin{tabular}{c c c c c c c}
Method & $\rho$ & $\bar\eta$ & $\bar\xi$& $\sigma^2_\eta$& $\sigma^2_\xi$&  $\rho$\\
&& Bias (s.d., s.e.) & Bias (s.d., s.e.) & Bias (s.d., s.e.) & Bias (s.d., s.e.)& Bias (s.d., s.e.) \\
\hline
{\it Logit link} &&&&&&\\
LIK & 0.2 & -0.195 (0.217, 0.213) &  -0.026 (0.150, 0.144) & -0.446 (0.312, 0.295) & -0.070 (0.151, 0.141) & 0.060 (0.189, 0.173) \\ 
MTM && 0.002 (0.261, 0.254) &  0.006 (0.157, 0.151) & -0.012 (0.473, 0.447) & -0.015 (0.172, 0.160) & 0.009 (0.248, 0.226)\\
LIK & 0.6 & -0.165 (0.216, 0.211) &  -0.013 (0.146, 0.142) & -0.444 (0.314, 0.297) & -0.075 (0.147, 0.136) & 0.142 (0.152, 0.140) \\ 
MTM && 0.014 (0.263, 0.252) & -0.006 (0.165, 0.150) & -0.007 (0.481, 0.454) & -0.017 (0.168, 0.160) & -0.007 (0.186, 0.165) \\
LIK & 0.8 & -0.157 (0.206, 0.208) &  -0.004 (0.140, 0.140) & -0.465 (0.302, 0.292) & -0.081 (0.151, 0.135) & 0.184 (0.126, 0.111) \\ 
MTM && 0.015 (0.257, 0.254) & -0.008 (0.155, 0.154) & -0.020 (0.480, 0.448) & -0.015 (0.191, 0.161) & -0.015 (0.143, 0.109) \\

{\it Probit link} &&&&&&\\
LIK & 0.2 & -0.063 (0.204, 0.201) &  0.010 (0.143, 0.138) & -0.249 (0.263, 0.243) & -0.034 (0.126, 0.114) & 0.025 (0.200, 0.177) \\ 
MTM && 0.004 (0.233, 0.237) & 0.002 (0.153, 0.150) & -0.014 (0.464, 0.446) & -0.008 (0.175, 0.162) & 0.000 (0.230, 0.207) \\
LIK & 0.6 & -0.070 (0.204, 0.203) &  0.010 (0.142, 0.139) & -0.230 (0.262, 0.241) & -0.027 (0.133, 0.116) & -0.073 (0.147, 0.131) \\ 
MTM && 0.005 (0.237, 0.241) & 0.005 (0.153, 0.151) & 0.024 (0.470, 0.445) & 0.003 (0.185, 0.165) & -0.004 (0.164, 0.146) \\
LIK & 0.8 & -0.073 (0.204, 0.202) &  0.012 (0.141, 0.138) & -0.246 (0.259, 0.238) & -0.036 (0.124, 0.112) & 0.102 (0.097, 0.090) \\ 
MTM && -0.002 (0.234, 0.236) & 0.006 (0.152, 0.149) & -0.015 (0.463, 0.422) & -0.020 (0.170, 0.157) & -0.003 (0.103, 0.092) \\

{\it Cloglog link} &&&&&&\\

LIK & 0.2 & -0.112 (0.171, 0.171) &  -0.077 (0.123, 0.119) & -0.598 (0.195, 0.181) & -0.163 (0.100, 0.093) & 0.040 (0.194, 0.170) \\ 
MTM && -0.001 (0.236, 0.240) & 0.006 (0.153, 0.152) & 0.001 (0.499, 0.458) & 0.005 (0.174, 0.171) & 0.003 (0.232, 0.204) \\
LIK & 0.6 & -0.092 (0.168, 0.167) & -0.083 (0.115, 0.118) & -0.619 (0.187, 0.178) & -0.173 (0.094, 0.088) & 0.116 (0.141, 0.128) \\ 
MTM && -0.092 (0.168, 0.167) & -0.083 (0.115, 0.118) & -0.619 (0.187, 0.178) & -0.173 (0.094, 0.088) & 0.116 (0.141, 0.128) \\
LIK & 0.8 & -0.103 (0.175, 0.168) &  -0.078 (0.119, 0.117) & -0.618 (0.179, 0.175) & -0.182 (0.095, 0.086) & 0.148 (0.098, 0.091) \\ 
MTM && -0.010 (0.246, 0.235) & 0.011 (0.153, 0.150) & -0.022 (0.474, 0.436) & -0.003 (0.174, 0.162) & 0.013 (0.109, 0.094) \\
 
\end{tabular}
\end{center}
%\label{tab:1}
\end{table}%

% low accuracy, n 25 , pi 35
\begin{table}[htp]
\caption{Bias (standard deviation s.d., average standard error s.e.) for the estimators of $\bar\eta, \bar\xi, \sigma^2_\eta, \sigma^2_\xi, \rho$ obtained from the approximate likelihood approach (LIK) and from the MTM approach (MTM), under increasing $\rho$ and different link function. Low accuracy of the test. Sample size $n=25$. Prevalence of disease $\pi=0.35$.}
\begin{center}
\begin{tabular}{c c c c c c c}
Method & $\rho$ & $\bar\eta$ & $\bar\xi$& $\sigma^2_\eta$& $\sigma^2_\xi$&  $\rho$\\
&& Bias (s.d., s.e.) & Bias (s.d., s.e.) & Bias (s.d., s.e.) & Bias (s.d., s.e.)& Bias (s.d., s.e.) \\
\hline
{\it Logit link} &&&&&&\\
LIK & 0.2 & -0.119 (0.217, 0.210) &  -0.032 (0.143, 0.146) & -0.324 (0.321, 0.298) & -0.075 (0.158, 0.142) & 0.046 (0.183, 0.172) \\ 
MTM && -0.003 (0.243, 0.237) &  0.012 (0.151, 0.154) & -0.039 (0.435, 0.396) & -0.009 (0.179, 0.164) & -0.004 (0.235, 0.219)\\
LIK & 0.6 & -0.091 (0.206, 0.209) &  -0.023 (0.143, 0.144) & -0.330 (0.332, 0.294) & -0.090 (0.149, 0.138) & 0.134 (0.159, 0.138) \\ 
MTM && 0.006 (0.237, 0.236) & 0.005 (0.155, 0.154) & -0.037 (0.445, 0.394) & -0.016 (0.178, 0.164) & 0.003 (0.190, 0.158) \\
LIK & 0.8 & -0.078 (0.218, 0.209) &  -0.018 (0.141, 0.142) & -0.303 (0.333, 0.302) & -0.097 (0.152, 0.137) & 0.157 (0.118, 0.105) \\ 
MTM && 0.008 (0.256, 0.236) & 0.000 (0.159, 0.152) & 0.008 (0.456, 0.411) & -0.013 (0.184, 0.172) & -0.013 (0.127, 0.102) \\
{\it Probit link} &&&&&&\\
LIK & 0.2 & -0.044 (0.215, 0.204) &  0.004 (0.141, 0.135) & -0.188 (0.265, 0.246) & -0.054 (0.119, 0.110) & 0.013 (0.194, 0.178) \\ 
MTM && -0.002 (0.238, 0.229) & -0.004 (0.157, 0.151) & -0.025 (0.423, 0.397) & -0.017 (0.181, 0.164) & -0.010 (0.218, 0.201) \\
LIK & 0.6 & -0.036 (0.207, 0.203) &  0.010 (0.134, 0.137) & -0.196 (0.260, 0.245) & -0.046 (0.120, 0.111) & 0.075 (0.150, 0.131) \\ 
MTM && 0.009 (0.227, 0.229) & 0.003 (0.145, 0.151) & -0.030 (0.405, 0.393) & -0.014 (0.173, 0.162) & 0.007 (0.165, 0.144) \\
LIK & 0.8 & -0.045 (0.209, 0.205) &  0.009 (0.142, 0.136) & -0.183 (0.270, 0.248) & -0.051 (0.118, 0.109) & 0.091 (0.100, 0.087) \\ 
MTM && 0.002 (0.226, 0.229) & 0.004 (0.156, 0.152) & -0.020 (0.423, 0.390) & -0.014 (0.169, 0.159) & 0.000 (0.109, 0.089) \\

{\it Cloglog link} &&&&&&\\

LIK & 0.2 & -0.098 (0.183, 0.177) &  -0.087 (0.122, 0.117) & -0.481 (0.204, 0.199) & -0.176 (0.100, 0.090) & 0.029 (0.190, 0.172) \\ 
MTM && 0.005 (0.235, 0.233) & 0.001 (0.159, 0.152) & -0.007 (0.436, 0.423) & 0.001 (0.183, 0.169) & -0.001 (0.224, 0.203) \\
LIK & 0.6 & -0.087 (0.175, 0.176) & -0.079 (0.119, 0.116) & -0.489 (0.206, 0.200) & -0.185 (0.096, 0.088) & 0.106 (0.142, 0.127) \\ 
MTM && 0.007 (0.230, 0.230) & 0.011 (0.152, 0.152) & -0.016 (0.468, 0.415) & -0.003 (0.183, 0.169) & 0.014 (0.165, 0.144) \\
LIK & 0.8 & -0.084 (0.178, 0.175) &  -0.085 (0.118, 0.116) & -0.490 (0.201, 0.196) & -0.183 (0.101, 0.090) & 0.123 (0.091, 0.085) \\ 
MTM && 0.007 (0.236, 0.229) & 0.006 (0.154, 0.151) & -0.027 (0.433, 0.395) & 0.000 (0.184, 0.166) & 0.002 (0.099, 0.086) \\

\end{tabular}
\end{center}
%\label{tab:1}
\end{table}%

\end{landscape}

%% figura low accuracy pi 20

\begin{figure}[htbp]
\begin{center}
\includegraphics[width=4.5in]{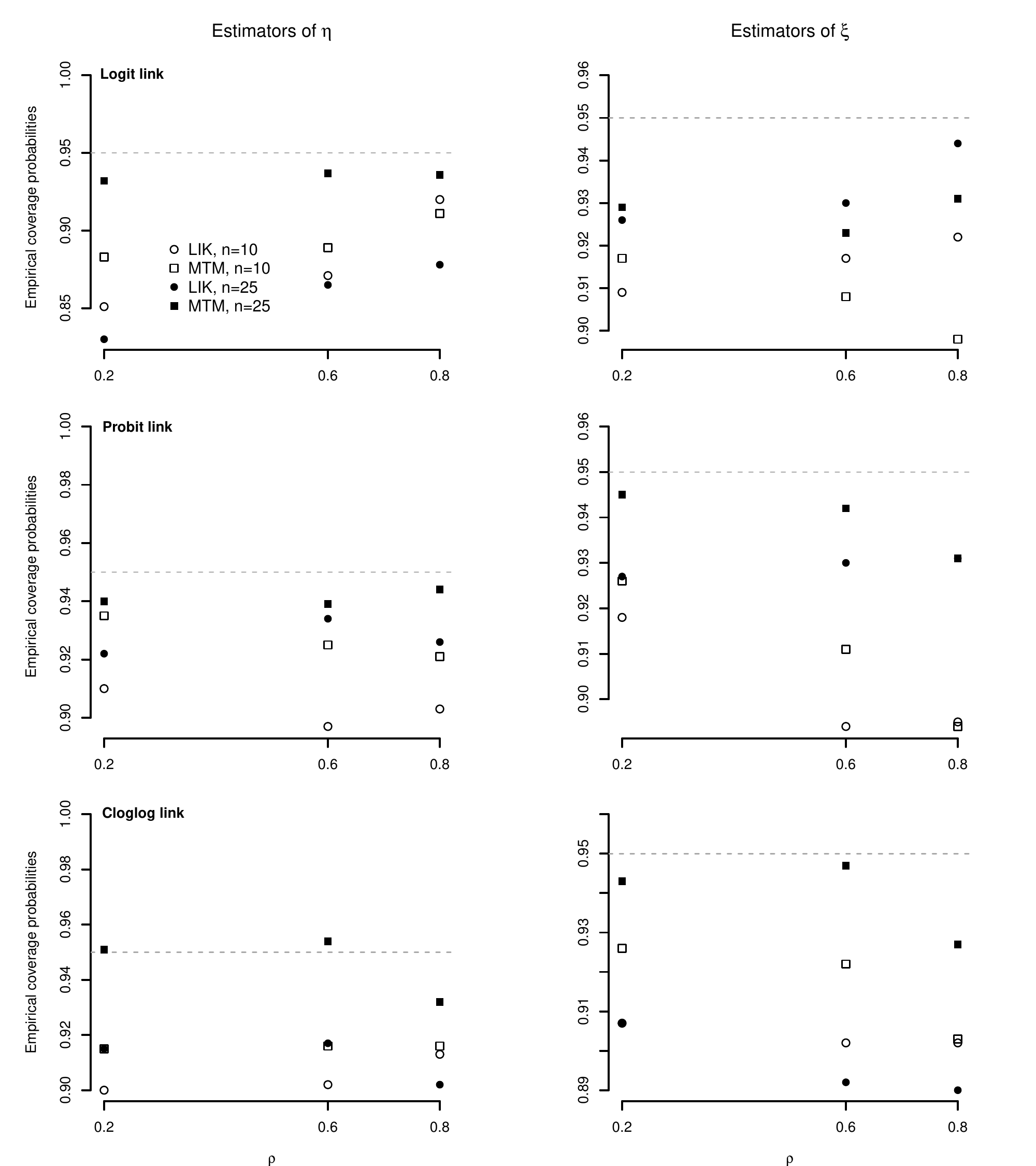}
\caption{{ Empirical coverage probability of Wald-type confidence interval for the estimators of $\bar\eta$ and $\bar\xi$ obtained from the approximate likelihood approach (LIK) and from the MTM approach (MTM), under increasing $\rho$, increasing sample size $n$ and different link function. Low accuracy of the test. Prevalence of disease $\pi=0.20$. Dashed line: nominal 95\% level.}}
\label{default}
\end{center}
\end{figure}

%% figura low accuracy pi 35

\begin{figure}[htbp]
\begin{center}
\includegraphics[width=4.5in]{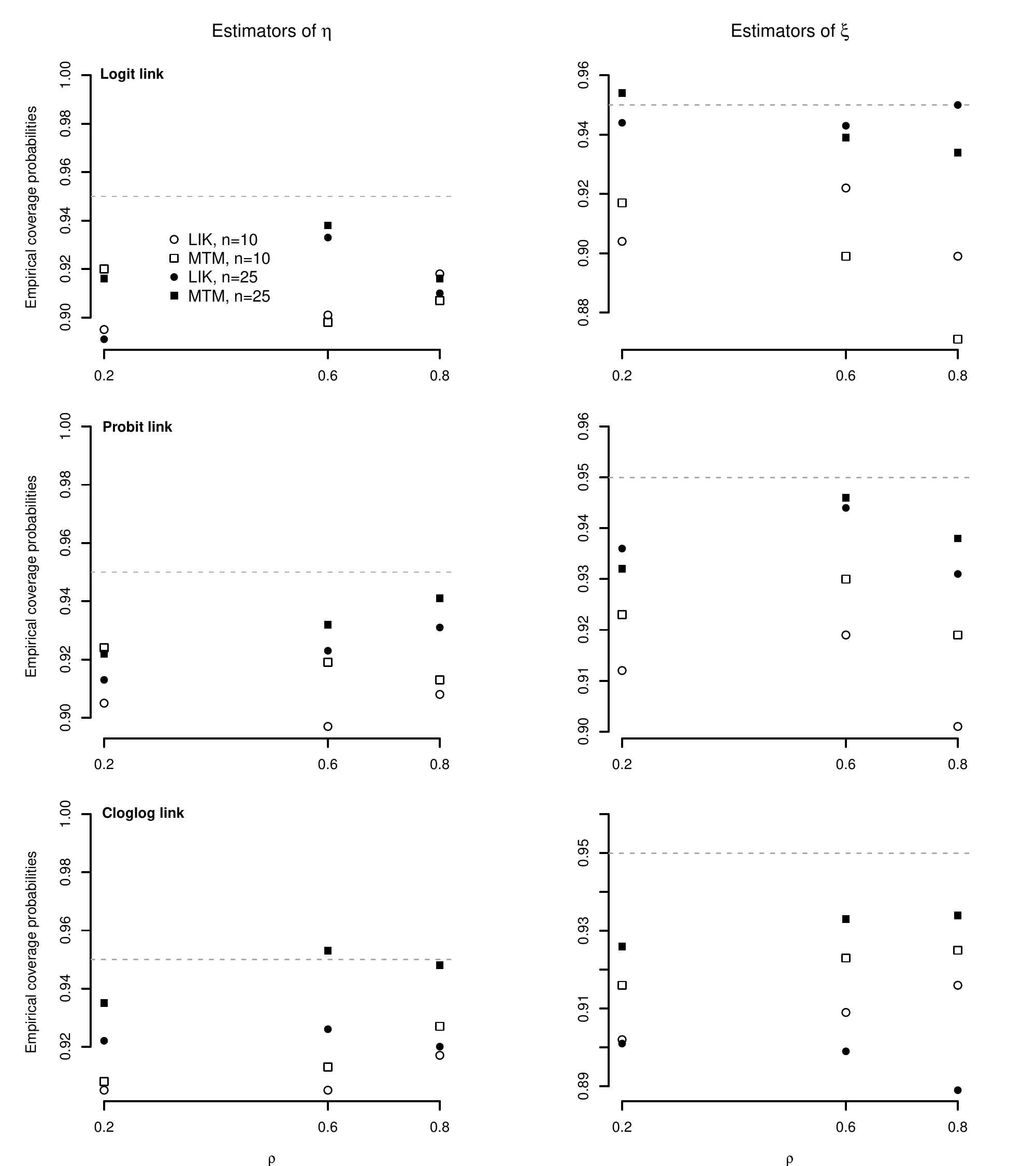}
\caption{{ Empirical coverage probability of Wald-type confidence interval for the estimators of $\bar\eta$ and $\bar\xi$ obtained from the approximate likelihood approach (LIK) and from the MTM approach (MTM), under increasing $\rho$, increasing sample size $n$ and different link function. Low accuracy of the test. Prevalence of disease $\pi=0.35$. Dashed line: nominal 95\% level.}}
\label{default}
\end{center}
\end{figure}

\end{document}